\documentclass[paper]{JHEP3}
\usepackage{epsfig,graphicx,bm,amsmath}

\title{S-branes and (Anti-)Bubbles in (A)dS Space}
\author{Dumitru Astefanesei\thanks{Alternative e-mail: {\tt dastefanesei@perimeterinstitute.ca}}\\
Harish-Chandra Research Institute, Chhatnag Road, Jhusi, Allahabad 211019, INDIA\\
E-mail: \email{dastef@mri.ernet.in}}
\author{Gregory C. Jones\\
Harvard University, Cambridge, MA 02138 USA\\
E-mail: \email{jones@physics.harvard.edu}}

\abstract{We describe the construction of new locally asymptotically 
(A)dS geometries with relevance for the AdS/CFT and
dS/CFT correspondences. Our approach is to obtain new
solutions by analytically continuing 
black hole solutions. A basic consideration of the
method of continuation indicates that these solutions come 
in three classes: S-branes, bubbles and anti-bubbles. A 
generalization to spinning or twisted solutions can yield
spacetimes with complicated horizon structures. Interestingly 
enough, several of these spacetimes are nonsingular.}
\preprint{HUTP-05/A0007,\ hep-th/0502162}

\begin{document}

\vfill \setcounter{page}{0} \setcounter{footnote}{0}
\newpage
\section{Introduction}
Time-dependent backgrounds in string theory provide an 
interesting arena for exploring intrinsically dynamical 
issues such as black hole evaporation, cosmological evolution 
or the possible formation and resolution of singularities. 
An essential ingredient in understanding quantum gravity in 
asymptotically Anti-de Sitter (AdS) spacetimes is the Maldacena 
conjecture (or the AdS/Conformal Field Theory (CFT) correspondence)
\cite{Maldacena:1997re,Witten:1998qj,Gubser:1998bc}.\footnote{It
is referred to as a duality 
in the sense that the supergravity (closed string) description 
of D-branes and the field theory (open string) description 
are different formulations of the same physics.} In this 
framework, a {\it large} black hole in AdS is described as a thermal 
state of the dual conformal field theory.  A remarkable 
property of the AdS/CFT correspondence is that it works 
even far from the conformal regime \cite{Boonstra:1998mp,Balasubramanian:1999jd}.
This result is consistent 
with the interpretation of the radial coordinate of AdS space 
as a energy scale of the dual CFT. In other words, timelike
D-branes lead to a spacelike holography.

Inspired by the fact that the microphysical statistical 
origin of cosmological horizon entropy may well be associated 
with a holographic dual theory, some authors conjectured a 
de Sitter/CFT correspondence \cite{Strominger:2001pn,Witten:2001kn,Balasubramanian:2001rb} 
--- the bulk time translation is dual to the boundary scale 
transformation and so the time is holographically reconstructed. 
Using the analogy with D-branes, one expects 
new (spacelike) objects S-branes to be at the heart of the 
dS/CFT correspondence. An S-brane \cite{Gutperle:2002ai}
is a topological defect 
for which time is a transverse dimension and so it exists 
only for a moment (or brief period) of time.
In the same way that (for $\Lambda=0$) $p$-branes are stationary solutions of
supergravity and string theory, S-branes are time-dependent backgrounds of the theory.

In this paper we find three families of exact solutions: S-branes, bubble-like solutions and the newly
coined anti-bubble solutions.  Roughly speaking, in $D$ dimensions, these solutions
involved a $(D-2)$-dimensional hyperbolic space, de Sitter, or anti-de Sitter component, respectively. The solutions are 
classified according to the technique of their construction.  (See also \cite{liu}.)

The first 
is the S-brane type
\cite{chen,Jones:2004rg,Wang:2004by,Tasinato:2004dy,Lu:2004ye,Gutperle:2004vh,
Jones:2004pz,cones}
describing a shell of radiation coming in from 
infinity and creating an unstable brane which subsequently 
decays.\footnote{The solutions in \cite{Jones:2004rg} do not have horizons and
are better described as gravitational wave solutions, describing the
creation and decay of a fluxbrane.  They were
constructed by analytic continuation keeping in mind Sen's rolling
tachyon solution for unstable D-branes \cite{Sen:2002nu}.} 
Nonspinning S-branes solutions involve ${\bf H}_{D-2}$ which
can be quotiented, to yield topological (A)dS black holes --- 
these have been known (see e.g. \cite{klemm,Cai:2001jd} and references 
therein). For example, a spherical black 
hole with $\Lambda>0$, under analytic continuation and sign flip of
metric, gives a black solution with $\Lambda<0$ and a hyperbolic
component.  We may refer to this as a (topological) black hole in AdS,
or as the corresponding S-brane solution to a black hole in dS.
On the other hand, a BHAdS with $\Lambda<0$ yields a cosmologically
singular S-BHAdS with $\Lambda>0$; this solution has an exterior
region \cite{carterleshouches} which is time-dependent, like de Sitter space
itself.\footnote{We emphasize a solution being time-dependent if it does
not have an exterior/asymptotic
region with a timelike Killing vector. Thus the Schwarzschild
black hole is not time-dependent even though there is no global timelike
Killing vector, and de Sitter space is time-dependent.}
The S-BH(A)dS solutions have timelike singularities
and the Penrose diagrams are 
related to a $\pi/2$-rotation of the corresponding black hole
Penrose diagrams \cite{klemm}. However, in the Reissner-Nordstr\o m case, the inmost horizon is moved to negative $r$ and the $r>0$ 
S-brane Penrose diagram has fewer regions.
The solutions we describe here are analogs of the S-branes found previously (with $\Lambda=0$) and that justifies the terminology.

With $\Lambda=0$, a black hole is stationary and an S-brane is
time-dependent, but a $\Lambda\neq 0$ will dominate at large $r$
and its sign determines the signature of the Killing vector.
Black holes and S-branes in AdS are both stationary,
and black holes and S-branes in dS are both time-dependent.

The second family are of bubble type \cite{Witten:1981gj,Aharony:2002cx,Birmingham:2002st,Balasubramanian:2002am,Biswas:2004xc}.
A bubble is a $(D-3)$-sphere
which exists only for $r\geq r_{\rm min}$.  An $x^D$ Killing circle
vanishes at $r=r_{\rm min}$.  These bubbles are time-dependent
since the $(D-3)$-sphere expands in a de Sitter fashion.  We also define
`double bubbles' as solutions where an expanding $(D-3)$-sphere exists
over an interval $r_{\rm min}\leq r\leq r_{\rm max}$ and the $x^D$-circle
closes at both endpoints (hence two `bubbles of nothing').

The third family is the newly coined anti-bubbles, which must be
distinguished from expanding bubbles.  Here, we have AdS$_{D-2}$
whose spatial section is not a sphere but a noncompact
`anti-bubble.'  This spatial section exists only for $r\geq r_{\rm min}$.
We also find double anti-bubbles where the AdS$_{D-2}$ runs over
$r_{\rm min}\leq r\leq r_{\rm max}$.

When rotation parameters are added (hence looking at Kerr-(A)dS solutions),
there is an additional complication
to the solution where a quantity $W$ (or $\Delta_\theta$) can vanish.
This can generate additional horizons changing the time-dependent or
stationary nature of various regions in the solution.  Also, sometimes
this will close the spacetime creating boundary conditions with inconsistent
Killing compactifications.

In some cases involving rotation, there are two types of S-branes.
For example, the $D=4$
Kerr solution admits a usual S-brane \cite{Wang:2004by} and also
a $\pi/2$-S-brane \cite{cones}.  This is analogous to the double
Killing Kerr bubble (with horizons and CTCs) as opposed to the $\theta\to\pi/2+i\theta$
Kerr bubble (without horizons or CTCs).
The idea is that in even dimensions, one direction
cosine is not associated with a rotation and hence it is different;
in any dimension, direction cosines with rotations turned off are
different from those with rotations turned on.
We will be careful to emphasize when such different solutions are available.

The rest of the paper is organized as follows: In Sec.~\ref{secondsec}
we look at the simple case
of RN(A)dS black holes in $D=4$ and find the bubbles, S-branes and
anti-bubbles (as well as interior double bubbles and anti-bubbles)
using card diagram\footnote{Card diagrams are applicable for $D=4$ or $5$ black
hole spacetimes which have the requisite $2$ or $3$ Killing fields.  Card
diagrams and the technique of the $\gamma$-flip were used to understand
S-branes in \cite{Jones:2004pz}.} techniques and 
using $r\theta$ diagrams.  Then in Sec.~\ref{thirdsec}
we look at the general-$D$
RN(A)dS solutions, finding bubbles, S-branes and anti-bubbles.
We see how the conformal boundary geometry of the S-brane fits nicely
with that for the bubble to give the global boundary of AdS.

In Sec.~\ref{fourthsec} we move to the Kerr solutions. These solutions are sometimes
plagued by what we call $W=0$ coordinate
singularities \cite{Gibbons:2004uw} (also called $\Delta_\theta=0$ singularities
\cite{Balasubramanian:2002am}).  We
find that these are just spinning Killing horizons (or twisted closures of
spatial Killing circles), which complicate
the structure of the spacetime. We allow general rotation parameters
and try to avoid $W=0$ singularities. Then, in Sec.~\ref{wzero}, we 
only turn one rotation on and
allow $W=0$ singularities. Here we find extremely interesting global
structures for bubble geometries with $W=0$ singularities and 
illustrate them by drawing skeleton diagrams\footnote{A skeleton 
diagram is a 1-dimensional analog of a card diagram; it shows only 
the coordinate which determines where the horizons are. The skeleton diagram for the Schwarzschild black hole is a $+$,
where the horizontal legs are where $r$ is spacelike and the vertical legs where
$r$ is timelike.  Four legs meet at a nonextremal horizon.} for the $\theta$ coordinate.

We conclude in Sec.~\ref{sixthsec} by outlining the role of these solutions in the holographic
AdS/CFT correspondence (or the putative dS/CFT correspondence).  Lastly we
give a short appendix on generalized card diagrams as
they apply to pure (A)dS$_D$ space for $D=4,5$.

We will not look at genus-zero planar or toroidal black holes or their
generalizations --- see \cite{klemm,Cai:2001jd}
and references contained therein.

\section{4d Examples}
\label{secondsec}

Before writing down the general analytic continuation, we look at
the simple case of four dimensions.  Here (as well as in five dimensions)
black holes in (A)dS are of Weyl type: in $D$ dimensions they have
$D-2$ commuting Killing fields.  Methods of obtaining bubbles, anti-bubbles
and S-branes are then very evident (see Figs.~\ref{RNdS4fig}-\ref{RNAdS4Qfig}).
Unlike previous approaches to analytic continuation to S-branes (involving
continuations like $r\to it$),
we will find all the spacetimes by only performing simple analytic continuations
involving real sections of hyperboloids, by making sign flips in the metric and sometimes continuing to imaginary charge.

We begin discussing the variety of solutions we will obtain starting with
the 4d Reissner-Nordstr\o m-(A)dS black hole solution.  From the 4d
RNdS black hole with $\Lambda>0$ we can obtain an S-brane
with $\Lambda<0$ as well as a static `anti-bubble' (which is to an expanding
bubble what AdS$_2$ is to dS$_2$) with $\Lambda<0$. From the 4d RNAdS
black hole $\Lambda<0$ we can obtain a bubble solution
with $\Lambda<0$ as well as an S-brane with $\Lambda>0$.  

\subsection{De Sitter}

\begin{figure}[htb]
\begin{center}
\epsfxsize=3in\leavevmode\epsfbox{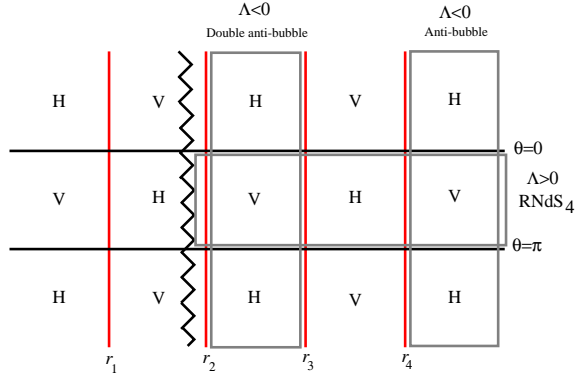}
\caption{Those regions of the extended RNdS$_4$ spacetime, where we
do not send $Q\to iQ$.  We have the black hole with $\Lambda>0$,
the anti-bubble with $\Lambda<0$, and the $r_2\leq r\leq r_3$
double anti-bubble with $\Lambda<0$.  This $r\theta$ diagram is
similar to the C-metric diagrams in \cite{bicak}.}
\label{RNdS4fig}
\end{center}
\end{figure}

\begin{figure}[htb]
\begin{center}
\epsfxsize=3in\leavevmode\epsfbox{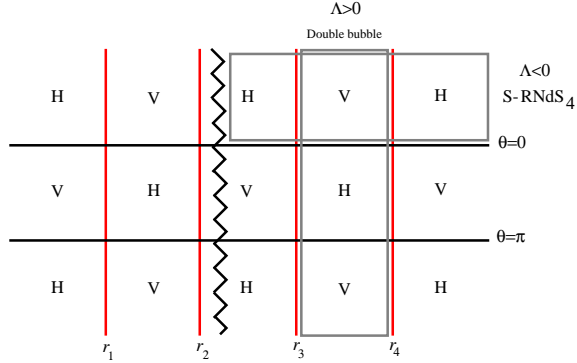}
\caption{Those regions of the extended RNdS$_4$ spacetime, where we
send $Q\to iQ$.  We have the $r_3\leq r\leq r_4$ double bubble
with $\Lambda>0$, and the S-RNdS$_4$ with $\Lambda<0$.}
\label{RNdS4Qfig}
\end{center}
\end{figure}

The RNdS$_4$ solution with $\Lambda=(D-1)/l^2$, $\Lambda>0$ is
\begin{eqnarray}\label{RNdS4sol}
ds^2&=&-f(r)(dx^4)^2+dr^2/f(r)+r^2d\Omega_2^2,\\
A&=&Q dx^4/r\nonumber
\end{eqnarray}
where $f(r)=1-2M/r+Q^2/r^2-r^2/l^2$.  We take the horizon function to be
the quartic polynomial $r^2f(r)=-r^4/l^2+r^2-2Mr+Q^2$, which can have
up to four roots\footnote{See \cite{klemm} for a discussion of roots and parameters.
However, the triple root $r_2=r_3=r_4$ is singular.}  $r_1<0<r_2\leq r_3\leq r_4$.  The root $r_4$ is the cosmological
horizon of de Sitter, and $r_3$ and $r_2$ are the outer and inner black
hole horizons, which can coincide in an `extremal' case.  The singularity is
at $r=0$, and $r<0$ with its single (cosmological)
horizon $r_1$ represents a negative-mass black hole in dS$_4$.  We can
draw the two non-Killing directions $r$, $\theta$ on the diagram in
Fig.~\ref{RNdS4fig}.  The RNdS$_4$ black hole solution occupies the middle
row, to the right of the singularity.
On this and also Fig.~\ref{RNdS4Qfig} we anticipate
several solutions that can be obtained from RNdS$_4$ by trivial analytic
continuation.

We can explicitly give $d\Omega_2^2=d\theta^2+\sin^2\theta d\phi^2$.
An alternative way to write it (to make contact with later formulas) is
to set $-1\leq\mu_0=\cos\theta\leq 1$ and $0\leq\mu_1=\sin\theta\leq 1$; we
have the constraint $\mu_0^2+\mu_1^2=1$.

This 4d spacetime has two commuting Killing vectors (say, the $x^4$ and
$\phi$ directions) and so is of generalized Weyl type.  The regions
in Fig.~\ref{RNdS4fig} are labelled H(orizontal) and V(ertical) in
analogy with Weyl card diagrams \cite{Jones:2004pz} (and see
Appendix A to this paper).  Horizontal
cards are stationary regions of spacetimes, whereas vertical cards
are time-dependent and have $D-2$ commuting spatial Killing fields.
There are two basic operations one can perform on cards.  On a horizontal
card, one can do a double Killing continuation to pick a new time
direction.  On vertical cards, one can perform an operation
known as a $\gamma$-flip which exchanges the timelike and spacelike
character of the two non-Killing directions.  The $\gamma$-flip
can be realized by changing the sign of the metric $g_{\mu\nu}\to -g_{\mu\nu}$
and analytically continuing all Killing directions; here
$x^4\to ix^4$ and $\phi\to i\phi$.  Along with the sign flip of the
metric, from the Einstein-Maxwell-$\Lambda$ equation we must flip
the sign of $\Lambda$.  Our conventions are to leave the parameter
$l$ alone but now interpret the solution to solve a $\Lambda<0$ equation.
The sign flip of the metric also forces the gauge field strength to
become imaginary; but we also continue the 1-form $dx^4\to idx^4$, so
the net result is a real field strength.  In summary, the $\gamma$-flip
takes a signature $(D-1,1)$ vertical card with a real field strength
and given $\Lambda$ and turns it `on its side' to yield 
a signature $(D-1,1)$ vertical card with a real
field strength and opposite sign of $\Lambda$.

It is clear then that we can take the vertical-card $r\geq r_4$ region
and turn the card on its side with a $\gamma$-flip.  We now occupy
the right column of Fig.~\ref{RNdS4fig}.  The solution is
\begin{eqnarray*}
ds^2&=&-f(r)(dx^4)^2-{dr^2\over f(r)}-r^2d\theta^2+r^2\sin^2\theta d\phi^2\\
&\rightarrow&-f(r)(dx^4)^2-{dr^2\over f(r)}+r^2 d\theta^2-r^2\sinh^2\theta d\phi^2.
\end{eqnarray*}
We decompactify $\phi$ to get horizons at $\theta=0,\pi$; the $0\leq\theta\leq\pi$
variable can be continued $\theta\to i\theta$, $\theta\to\pi+i\theta$ to
give a patched representation of AdS$_2$.  We  must compactify
$x^4\simeq x^4+4\pi |f'(r_4)|^{-1}$ to avoid a conical singularity
at $r=r_4$.  In summary, we have AdS$_2$ and an $x^4$-circle fibered
over $r\geq r_4$.  At $r=r_4$ the $x^4$-circle closes in a fashion
very similar to well-known expanding bubble solutions giving a minimum-size
AdS$_2$.  In analogy with `bubble' terminology we shall call this solution \cite{Birmingham:2002st} the RNdS$_4$ anti-bubble, with $\Lambda<0$.

Note that we could also perform the $\gamma$-flip on $r_2\leq r\leq r_3$. Now the spacetime has two boundaries $r=r_2,r_4$ where the $x^4$-circle closes. We must then match $f'(r_3)+f'(r_4)=0$ to eliminate conical singularities
at both ends; then of the parameters $M$, $Q$ and $l$, one is dependent --- one can also be eliminated by a global conformal transformation, leaving
one true dimensionless shape parameter. This solution occupies the
center column in Fig.~\ref{RNdS4fig}, and we call it the $r_2\leq r\leq r_3$
RNdS$_4$ double anti-bubble.  Since it does not have an $r\to\infty$ asymptotic
region, it is not useful for holography.

The region $r_3\leq r\leq r_4$ is a stationary region (horizontal card)
and we may perform a double Killing continuation $x^4\to ix^4$, $\phi\to i\phi$
to get a new solution.  We must also continue $Q\to iQ$ to make the field
strength real.  Then the horizon function $r^2f(r)=1-2M/r-Q^2/r^2+r^2/l^2$
is changed and its roots are generically $r_1\leq r_2<0<r_3\leq r_4$.  We
now reference solutions to Fig.~\ref{RNdS4Qfig}; note that positive-
and negative-mass solutions are qualitatively similar.  The solution is
\begin{eqnarray}
ds^2&=&f(r)(dx^4)^2+{dr^2\over f(r)}+r^2(d\theta^2-\sin^2d\phi^2)\\
&\rightarrow&f(r)(dx^4)^2+{dr^2\over f(r)}+r^2(-d\theta^2+\sinh^2\theta d\phi^2).\nonumber
\end{eqnarray}
We see now a patched dS$_2$; and the $x^4$ circle vanishes at
$r=r_3,r_4$, so we require $f'(r_3)+f'(r_4)=0$ and have
the $r_3\leq r\leq r_4$ RNdS$_4$ double bubble.  It solves $\Lambda>0$
Einstein-Maxwell-$\Lambda$.  Since it does not
have an $r\to\infty$ asymptotic region, it is not useful for holography.

We can however take the vertical card at hyperbolic $\theta$ on the RNdS$_4$
double bubble and perform a $\gamma$-flip.  The resulting solution has
$\Lambda<0$ and we call it S-RNdS$_4$, since it is the S-brane gotten from
the RNdS$_4$ geometry.  It occupies the top row of Fig.~\ref{RNdS4Qfig},
to the right of the singularity.  It is
$$ds^2=f(r)(dx^4)^2-{dr^2\over f(r)}+r^2(d\theta^2+\sinh^2\theta d\phi^2).$$
We see ${\bf H}_2$, azimuthally parametrized.  Note that just as the RNdS$_4$
black hole is not stationary, its S-brane is not time-dependent.

The RNdS$_4$, RNdS$_4$ anti-bubble and S-RNdS$_4$ all have $r\to\infty$
asymptotic regions where they are locally asymptotic to (A)dS$_4$, depending
on their sign of $\Lambda$.

We summarize the five spacetimes gotten from RNdS$_4$ in the following table.
$$\begin{array}{ccccccc}
{\rm Name}&\Lambda&{\rm Hyp.}&iQ?&\phi{\rm\ cpct}&x^4{\rm\ cpct}&{\rm Asym.}\\
{\rm RNdS}_4&+&S^2&{\rm No}&{\rm Yes}&{\rm No}&{\rm dS}_4\\
{\rm RNdS}_4\,\mbox{anti-bub.}&-&{\rm AdS}_2&{\rm No}&{\rm No}&{\rm Yes}&{\rm AdS}_4\\
{\rm RNdS}_4\,{\rm doub.~bub.}&+&{\rm dS}_2&{\rm Yes}&{\rm No}&{\rm Double}&~\\
{\rm RNdS}_4\,\mbox{doub.~anti-bub.}&-&{\rm AdS}_2&{\rm No}&{\rm No}&{\rm Double}&~\\
\mbox{S-RNdS}_4&-&{\bf H}_2&{\rm Yes}&{\rm Yes}&{\rm No}&{\rm AdS}_4
\end{array}$$
Here we give the name, the sign of the cosmological constant, the real
section of the complex 2-hyperboloid,
whether $Q$ has been continued with $Q\to iQ$,
whether $\phi$ is compact, whether $x^4$ is compact, whether $x^4$ has
two boundaries instead of just one since this is a nontrivial 
condition and whether the manifold asymptotes locally to (A)dS$_4$.
The isometry group of the solution
is the isometry of the hyperbolic space ($SO(3)$, $SO(2,1)$ or $SO(1,2)$)
times ${\bf R}$ (if $x^4$ is noncompact)
or $U(1)$ (if $x^4$ is compact).

\subsection{Anti-de Sitter}

\begin{figure}[htb]
\begin{center}
\epsfxsize=2in\leavevmode\epsfbox{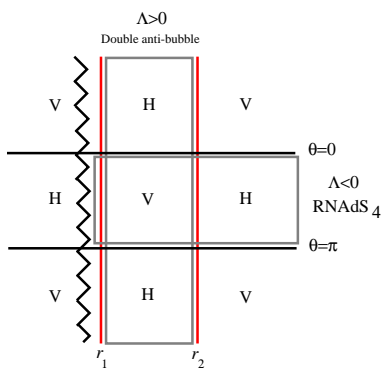}
\caption{Those regions of the extended RNAdS$_4$ spacetime, where we
do not send $Q\to iQ$.}
\label{RNAdS4fig}
\end{center}
\end{figure}

\begin{figure}[htb]
\begin{center}
\epsfxsize=2in\leavevmode\epsfbox{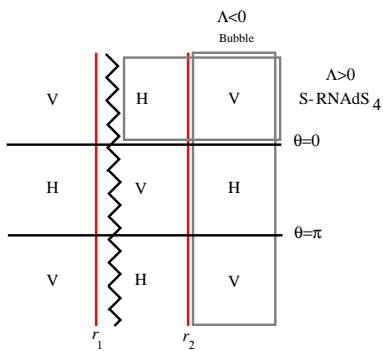}
\caption{Those regions of the extended RNAdS$_4$ spacetime, where we
send $Q\to iQ$.}
\label{RNAdS4Qfig}
\end{center}
\end{figure}

To achieve RNAdS$_4$, take $l\to il$ in (\ref{RNdS4sol}).  We have
\begin{eqnarray}
ds^2&=&-f(r)(dx^4)^2+dr^2/f(r)+r^2d\Omega_2^2,\\
A&=&Q dx^4/r\nonumber
\end{eqnarray}
where $f(r)=1-2M/r+Q^2/r^2+r^2/l^2$.  Then $r^2f(r)$ can have at
most two roots; we assume $Q\neq 0$ and $M$ is large enough so
that this happens. Then $0<r_1\leq r_2$ --- see Fig~\ref{RNAdS4fig}. 
Looking ahead, when we will have to send $Q\to iQ$, we will 
have $r_1<0<r_2$ --- see Fig.~\ref{RNAdS4Qfig}.

Since our discovery of solutions parallels the RNdS$_4$ case, we will be
brief.  We can take $r\geq r_2$ which is static, and double Killing
rotate $x^4\to ix^4$, $\phi\to i\phi$, $Q\to iQ$.  We then get the RNAdS$_4$ bubble solution with
a patched description of dS$_2$ fibered over $r\geq r_2$, and an $x^4$-circle which
closes at $r=r_2$ to give the bubble.  This is the right column of
Fig.~\ref{RNAdS4Qfig}.

We can perform a $\gamma$-flip on the upper (vertical card) region of the RNAdS$_4$
bubble to achieve S-RNAdS$_4$, which is the top row of Fig.~\ref{RNAdS4Qfig},
to the right of the singularity.  It has azimuthally parametrized ${\bf H}_2$ and
$\Lambda<0$.

Finally we can perform a $\gamma$-flip on $r_1\leq r\leq r_2$ of the RNAdS$_4$
black hole to achieve a double anti-bubble.  It has
AdS$_2$, $\Lambda>0$, and appears as the central column in Fig.~\ref{RNAdS4fig}.

The RNAdS$_4$ black hole, bubble and S-brane asymptote to (A)dS$_4$ locally.

Again we summarize
$$\begin{array}{cccccccc}
{\rm Name}&\Lambda&{\rm Hyp.}&iQ?&\phi{\rm\ cmpct}&x^4{\rm\ cmpct}&{\rm Asym.}\\
{\rm RNAdS}_4&-&S^2&{\rm No}&{\rm Yes}&{\rm No}&{\rm AdS}_4\\
{\rm RNAdS}_4\,{\rm bub.}&-&{\rm dS}_2&{\rm Yes}&{\rm No}&{\rm Yes}&{\rm AdS}_4\\
{\rm RNAdS}_4\,\mbox{doub.~anti-bub.}&+&{\rm AdS}_2&{\rm No}&{\rm No}&{\rm Double}&~\\
\mbox{S-RNAdS}_4&+&{\bf H}_2&{\rm Yes}&{\rm Yes}&{\rm No}&{\rm dS}_4
\end{array}$$

The fact that all the listed solutions are different from each other is
evident just by looking at where they stand in relation to their neighbors
and the singularity, in Figs.~\ref{RNdS4fig}-\ref{RNAdS4Qfig}. One can
also use the symmetry groups to prove they are different.

These 4d solutions with $SO(3)$ symmetry can also yield bubbles, anti-bubbles or S-branes based on the continuation $\theta\to \pi/2+i\theta$ instead
of $\theta\to i\theta$.  These solutions are not different from those solutions
just described.  However, even-dimensional Kerr solutions admit
$\theta\to\pi/2+i\theta$ solutions which are different from those gotten by
taking the analog of $\theta\to i\theta$.  This distinction has been emphasized
in \cite{Jones:2004pz,cones} and we will revisit it below.

The card-diagram method of the $\gamma$-flip also applies in 5d.
Card diagrams of (A)dS$_4$ and (A)dS$_5$ spacetime can
be drawn; in fact due to their extra symmetries, many different card diagram
representations exist.  These diagrams are also useful for visualizing the local-(A)dS$_4$, AdS$_5$ asymptotia of the black hole, bubble, anti-bubble,
and S-brane solutions.  For some details and diagrams,
see Appendix A to this paper.

However, card diagram methods do not apply in $D\geq 6$.  Nonetheless,
analogs of these RN solutions do exist in all $D\geq 4$, and we give them
in the next section.

\section{General Reissner-Nordstr\o m-(A)dS$_D$ Solutions}
\label{thirdsec}

We construct the S-branes for the general RN(A)dS$_D$ solution, along
with the bubbles and anti-bubbles.  We will not have
Figs.~\ref{RNdS4fig}-\ref{RNAdS4Qfig} to guide us,
but again we will only do simple analytic
continuations involving cos(h)-type quantities, metric sign flips, and
$Q\to iQ$.  We will not focus on double (anti-)bubble solutions, only
on those solutions with $r\to\infty$ asymptotia.  We also give the
$r\to\infty$ conformal boundary geometry (CBG) gotten from the given coordinates.

\subsection{De Sitter}

The RNdS$_D$ black hole is
\begin{eqnarray}
ds^2&=&-f(r)(dx^D)^2+{dr^2\over f(r)}+r^2 d\Omega_{D-2}^2,\\
A&=&\sqrt{D-2\over 2(D-3)}{Q dx^4\over r}\nonumber,
\end{eqnarray}
where $f(r)=1-2M/r^{D-3}+Q^2/r^{2(D-3)}-r^2/l^2$, $\Lambda=(D-1)/l^2>0$.
The horizon function is the polynomial $r^{2(D-3)}f(r)$.  For
$D$ even, an appropriate parameter subdomain gives
roots $r_1<0<r_2\leq r_3\leq r_4$ --- we
consider the solution for $r>0$.  For $D$ odd we let $r^2$
be the independent variable and allow it to go negative.  Then
for the appropriate parameter subdomain the horizon function has roots
which we call $0<r_2^2\leq r_3^2\leq r_4^2$
and we consider $r^2>0$ 
(there is no $r_1^2$ root in our notation).

Take $D=2n+2$ even first; we can write
\begin{equation}\label{evenspheremetric}
d\Omega_{D-2}^2=d\mu_0^2+d\mu_1^2+\cdots+
d\mu_n^2+\mu_1^2d\phi_1^2+\cdots+\mu_n^2d\phi_n^2,
\end{equation}
where $-1\leq\mu_0\leq 1$ and $0\leq\mu_i\leq 1$ $i=1,\ldots,n$.
The constraint is $\mu_0^2+\sum_{i=1}^n\mu_i^2=1$.

To get the S-brane, we send $\mu_i\to i\mu_i$, $i=1,\ldots,n$,
send $g_{\mu\nu}\to-g_{\mu\nu}$, and $Q\to iQ$.  The line element
(\ref{evenspheremetric}) becomes, including the sign flip, the (still spacelike)
\begin{equation}
d{\bf H}_{D-2}^2=-d\mu_0^2+d\mu_1^2+\cdots+
d\mu_n^2+\mu_1^2d\phi_1^2+\cdots+\mu_n^2d\phi_n^2
\end{equation}
with constraint $\mu_0^2-\sum_{i=1}^n\mu_i^2=1$.  The solution is
\begin{eqnarray*}
ds^2&=&f(r)(dx^D)^2-{dr^2\over f(r)}+r^2 d{\bf H}_{D-2}^2,\\
A&=&\sqrt{D-2\over 2(D-3)}{Q dx^4\over r}\nonumber,
\end{eqnarray*}
which has $\Lambda<0$.  As $r\to\infty$, the solution is asymptotically
locally AdS$_D$.  The conformal boundary geometry (CBG) in these coordinates is
$ds^2=-(dx^D)^2/l^2+d{\bf H}_{D-2}^2$, which is
${\bf R}_{\rm time}\times {\bf H}_{D-2}$.  There is no invariant
relating the size of these two components.  The horizon function now has
roots $r_1\leq r_2<0<r_3\leq r_4$ like the S-RNdS$_4$ case. Our 
$r\to\infty$ gets the asymptotia gotten from one Rindler wedge of 
the $r_4$ horizon --- there is another Rindler wedge ignored in this procedure.

Taking the S-brane, we can take $x^D\to ix^D$, continue back $Q\to -iQ$,
so $r_1<0<r_2\leq r_3\leq r_4$.  We compactify $x^D\simeq x^D+4\pi|f'(r_4)|^{-1}$
to form an anti-bubble
at $r=r_4$, and then take for example $\phi_1\to i\phi_1$.  Thus
$$ds^2=-f(r)(dx^D)^2-{dr^2\over f(r)}+r^2d{\rm AdS}_{D-2}^2.$$
Here,
$$d{\rm AdS}_{D-2}^2=-d\mu_0^2+d\mu_1^2+\cdots+d\mu_n^2-\mu_1^2d\phi_1^2
+\mu_2^2d\phi_2^2+\cdots+\mu_n^2d\phi_n^2$$
is a patched description; we can go through the $\mu_1$ Rindler
horizon to $\mu_1\to i\mu_1$.
This anti-bubble solution was discovered in planar coordinates
in \cite{Lu:2003iv} where they were termed fluxbranes.
In planar coordinates the solutions resemble branes but
keeping in mind their global structure, we choose not to think
of them as branes, the term anti-bubble (for the AdS$_{D-2}$ factor)
being more appropriate.

The CBG is $ds^2\propto (dx^D)^2/l^2+d{\rm AdS}_{D-2}^2$,
which is $S^1\times$AdS$_{D-2}$.
Since $x^D$ is compact, there is a dimensionless invariant, the ratio of
the circumference of the $x^D$-circle to the unit-sized AdS$_{D-2}$.

In odd dimension $D=2n+1$ the idea is the same, but the cosines are
set up differently.  We have $\mu_i$, $i=1,\ldots,n$, with $0\leq\mu_i\leq 1$ and
\begin{equation}\label{oddspheremetric}
d\Omega_{D-2}^2=d\mu_1^2+\cdots+d\mu_n^2+\mu_1^2d\phi_1^2+\cdots
+\mu_n^2d\phi_n^2.
\end{equation}
To get the S-brane, take $\mu_i\to i\mu_i$ $i=1,\ldots,n-1$,
$g_{\mu\nu}\to -g_{\mu\nu}$, $Q\to iQ$, flip the sign of $\Lambda$,
and $\phi_n\to i\phi_n$. 
Then we have \begin{equation}\label{oddhyper}
d{\bf H}_{D-2}^2=d\mu_1^2+\cdots+d\mu_{n-1}^2-d\mu_n^2+\mu_1^2d\phi_1^2+\cdots
+\mu_n^2d\phi_n^2,
\end{equation}
and the geometry is
$$ds^2=f(r)(dx^D)^2-{dr^2\over f(r)}+r^2 d{\bf H}_{D-2}^2.$$
The horizon function now has roots $r_2^2<0<r_3^2\leq r_4^2$.

To go to the anti-bubble, send $x^D\to ix^D$, return $Q\to -iQ$, and
send say $\phi_1\to i\phi_1$.  Then
$$d{\rm AdS}_{D-2}^2=d\mu_1^2+\cdots+d\mu_{n-1}^2-d\mu_n^2
-\mu_1^2d\phi_1^2+\mu_2^2d\phi_2^2+\cdots+\mu_n^2d\phi_n^2.$$
This is in the `real' $\mu_1$ patch where the constraint reads
$\mu_n^2-\mu_1^2-\mu_2^2-\cdots-\mu_{n-1}^2=1$, but going
through the Rindler horizon at $\mu_1=0$, we send $\mu_1\to i\mu_1$
and get $\mu_n^2+\mu_1^2-\mu_2^2-\cdots-\mu_{n-1}^2=1$.  The anti-bubble
is $$ds^2=-f(r)(dx^D)^2-{dr^2\over f(r)}+r^2 d{\rm AdS}_{D-2}^2.$$

The CBG is $ds^2\propto (dx^D)^2/l^2+d{\rm AdS}_{D-2}^2$,
which is $S^1\times$AdS$_{D-2}$.
Since $x^D$ was compactified at the largest $r$-root, there is
a dimensionless invariant, the circumference of the $x^D$-circle over
the unit AdS size.

\subsection{Anti-de Sitter}

{}From the RNAdS$_D$ solution we will make a $\Lambda<0$ bubble and
$\Lambda>0$ S-brane.  The solution is like for RNdS$_D$ except
$f(r)=1-2M/r^{D-3}+Q^2/r^{2(D-3)}+r^2/l^2$, with roots $0<r_1\leq r_2$.

For $D=2n+2$, $d\Omega_{D-2}^2$ is as in (\ref{evenspheremetric}).
To get the bubble, send $x^D\to ix^D$, compactify at $r_2$, and
send say $\phi_1\to i\phi_1$.  The solution is
\begin{equation}\label{RNAdSbubble}
ds^2=f(r)(dx^D)^2+{dr^2\over f(r)}+d{\rm dS}_{D-2}^2,
\end{equation}
where
$$d{\rm dS}_{D-2}^2=d\mu_0^2+d\mu_1^2+\cdots+d\mu_n^2-\mu_1^2d\phi_1^2
+\mu_2^2d\phi_2^2+\cdots+\mu_n^2d\phi_n^2.$$
Of course we can go through the Rindler horizon at $\mu_1=0$ and
send $\mu_1\to i\mu_1$.  This bubble was described in
\cite{Birmingham:2002st,Balasubramanian:2002am}.

The CBG is
$ds^2\propto (dx^D)^2/l^2+d{\rm dS}_{D-2}^2$
which is $S^1\times$dS$_{D-2}$.  There is a dimensionless invariant,
the ratio of the sizes of these factors.

To get the S-brane from the black hole, send $\mu_i\to i\mu_i$, $i=1,\ldots,n$,
$g_{\mu\nu}\to-g_{\mu\nu}$, $Q\to iQ$, and flip the sign of $\Lambda$.
Now $r_1<0<r_2$; the solution is
\begin{equation}\label{SRNAdS}
ds^2=f(r)(dx^D)^2-{dr^2\over f(r)}+r^2d{\bf H}_{D-2}^2,
\end{equation}
where
$$d{\bf H}_{D-2}^2=-d\mu_0^2+d\mu_1^2+\cdots+d\mu_n^2+\mu_1^2d\phi_1^2
+\cdots+d\mu_n^2.$$
The constraint is $\mu_0^2-\sum_{i=1}^n\mu_i^2=1$.  Since the
singularity is just protected by a cosmological horizon $r_2$,
this solution is nakedly singular, like the S-Schwarzschild geometry
of pure Einstein theory.

The CBG is $ds^2\propto (dx^D)^2/l^2+d{\bf H}_{D-2}^2$,
which is ${\bf R}_{\space}\times {\bf H}_{D-2}$.
This is a Euclidean geometry, so this would serve to investigate
the putative dS/CFT correspondence.

In odd dimensions $D=2n+1$, we have $d\Omega_{D-2}^2$ given by
(\ref{oddspheremetric}). The bubble is gotten by sending
$x^D\to ix^D$, $Q\to iQ$ and say $\phi_1\to i\phi_1$.  The
solution is (\ref{RNAdSbubble}), where
$$d{\rm dS}_{D-2}^2=
d\mu_1^2+\cdots+d\mu_n^2-\mu_1^2d\phi_1^2
+\mu_2^2d\phi_2^2+\cdots+\mu_n^2d\phi_n^2.$$

The odd-dimensional S-brane on the other hand is gotten from
$\mu_i\to i\mu_i$ $i=1,\ldots,n-1$, $g_{\mu\nu}\to-g_{\mu\nu}$,
$\phi_n\to i\phi_n$.  The solution is (\ref{SRNAdS})
with the hyperbolic space given by (\ref{oddhyper}).

\subsection{Extremal S-RNdS$_D$}
Take S-RNdS$_D$,
\begin{eqnarray}
ds^2&=&f(r)(dx^D)^2-dr^2/f(r)+r^2d{\bf H}_{D-2}^2\\
A&=&\sqrt{D-2\over 2(D-3)}{Qdx^D\over r}.
\end{eqnarray}
Here, $f(r)=1-2M/r^{D-3}-Q^2/r^{2(D-3)}-r^2/l^2$.
For $D$ even we normally assume four roots $r_1\leq r_2<0<r_3\leq r_4$.
For $D$ odd we have $r_2^2<0<r_3^2\leq r_4^2$. In
either case, one can find an extremal solution where $r_3=r_4$.
Here, `extremal' refers just to degenerate horizons; this
solution is the analog of the $r_3=r_4$ maximal RNdS$_D$ black
hole solution where the outer black hole horizon coincides with
the cosmological (de Sitter) horizon.
Then $f(r)\sim -A(r-r_4)^2$, and letting $\epsilon=r-r_4$,
we can take a scaling limit where $\epsilon\to 0$, $\epsilon x^D$
fixed, which is
$$ds^2=-A\epsilon^2(dx^D)^2+{d\epsilon^2\over A\epsilon^2}
+r_4^2d{\bf H}_{D-2}^2,$$
and $F\propto {Q\over r_4^{D-2}}dx^D\wedge d\epsilon$.  This
solution is AdS$_2\times {\bf H}_{D-2}$.  Thus extremal
S-RNdS$_D$ interpolates between AdS$_2\times {\bf H}_{D-2}$ at
the extremal horizon to local AdS$_D$ at $r=\infty$ and
the latter can have a CBG of
${\bf H}_{D-2}\times {\bf R}_{\rm time}$.

Solutions which interpolate between spacetimes with similar
${\bf H}_{D-2}$ factors were found in \cite{acharya}.  For de
Sitter bounces, see \cite{Lu:2004fe}.

\subsection{Embedding the Conformal Boundary Geometry of Bubbles and S-branes}

\begin{figure}[htb]
\begin{center}
\epsfxsize=1.5in\leavevmode\epsfbox{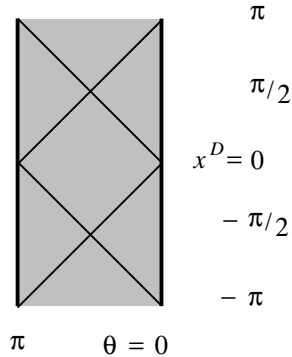}
\caption{Penrose diagram for the global conformal boundary of AdS$_D$.
The top and bottom $x^D=\pm\pi$ are identified.
Each diamond is filled in by the asymptotia of an
RNAdS$_D$ bubble, and S-RNdS$_D$ gives
triangles.  Four triangles and two diamonds neatly fit together
so their closure gives the whole global boundary.}
\label{asympfig}
\end{center}
\end{figure}

In \cite{Balasubramanian:2002am} the conformal boundary of the RNAdS$_D$
bubble was given as a subset of $S^1_{\rm time}\times S^{D-2}$ which is
the global conformal boundary of AdS$_D$ --- we have identified the time-circle
in the canonical fashion. There, it was found that in the $x^D\theta$ strip,
where $0\leq\theta\leq\pi$ is the polar angle for the $S^{D-2}$ of RNAdS$_D$, the
bubble asymptotes to the open set $|\theta-\pi/2|<|x^D-\pi/2|$; each bubble
asymptotes to one diamond in Fig.~\ref{asympfig}.  We now complete the picture
by showing that S-RNdS$_D$ asymptotes to the remainder triangles.

First, take $D=2n+1$ odd.  The embedding of dS$_D$ into ${\bf R}^{D,1}$
is
\begin{eqnarray*}
X'^0&=&\sqrt{1-r^2}\sinh x^D\\
X^1&=&r\mu_1\cos\phi_1\\
X^2&=&r\mu_1\sin\phi_1\\
&\vdots&\\
X^D&=&\pm\sqrt{1-r^2}\cosh x^D.
\end{eqnarray*}
Here, a prime denotes a timelike coordinate.
We want to send $\mu_i\to i\mu_i$ for $i=1,\ldots,n-1$,
$\phi_n\to i\phi_n$, and flip the sign of $g_{\mu\nu}$.
Then we get, upon also taking $r>1$,
\begin{eqnarray*}
X'^0&=&\sqrt{r^2-1}\sinh x^D\\
X^1&=&r\mu_1\cos\phi_1\\
X^2&=&r\mu_1\sin\phi_1\\
&\vdots&\\
X'^{D-2}&=&r\mu_n\cosh \phi_n\\
X^{D-1}&=&r\mu_n\sinh\phi_n\\
X^D&=&\pm\sqrt{r^2-1}\cosh x^D.
\end{eqnarray*}
Then $X'^{D-2}>0$ and we let $X'^0/X'^{D-2}=T$.  The
global-time angle $\tan^{-1}T$ runs from $-\pi/2$ to $\pi/2$.
Then
$\tan|\theta-\pi/2|=|X^D|/\sqrt{(X^{D-1})^2+(X^1)^2+(X^2)^2+\cdots}
=(\cosh x^d)/(\mu_n \cosh \phi_n)>|T|$,
so we have precisely two triangles from this description.

For $D=2n+2$ even, we have
\begin{eqnarray*}
X'^0&=&\sqrt{1-r^2}\sinh x^D\\
X^1&=&r\mu_1\cos\phi_1\\
X^2&=&r\mu_1\sin\phi_1\\
&\vdots&\\
X^{D-1}&=&r\mu_0\\
X^D&=&\pm\sqrt{1-r^2}\cosh x^D.
\end{eqnarray*}
Sending $\mu_i\to i\mu_i$ for $i=1,\ldots,n$ and flipping the
sign of $g_{\mu\nu}$, and going to $r>1$, we get
\begin{eqnarray*}
X'^0&=&\sqrt{r^2-1}\sinh x^D\\
X^1&=&r\mu_1\cos\phi_1\\
X^2&=&r\mu_1\sin\phi_1\\
&\vdots&\\
X'^{D-1}&=&r\mu_0\\
X^D&=&\pm\sqrt{r^2-1}\cosh x^D.
\end{eqnarray*}
We have $X'^{D-1}>0$ so set $X'^0/X'^{D-1}=(\sinh x^D)/\mu_0=T$,
and $-\infty<T<\infty$.  Furthermore, $|X^D|/\sqrt{(X^1)^2+(X^2)^2+\cdots}
=(\cosh x^D)/\mu_0>|T|$.

\section{Kerr-(A)dS$_D$ And Related Solutions which avoid $W=0$}
\label{fourthsec}

We now find S-branes, bubbles and anti-bubbles from the Kerr-(A)dS solutions.
In the notation of \cite{Gibbons:2004uw}, for bubbles, S-branes and anti-bubbles,
a quantity $W$ defined below has the possibility to zero along certain
hypersurfaces which are $r$-independent --- they depend on the cosines
and hyperbolic functions.

In this section, we will look for solutions which avoid
$W=0$, which are clearly good solutions, with an expected global structure.
The $W=0$ coordinate singularity for
the Kerr-AdS$_4$ bubble was a source of some confusion
in \cite{Balasubramanian:2002am}; actually it is a regular spinning horizon
with a constant angular velocity.  Following their approach, in Sec.~\ref{wzero}
we will look at the case of general $D$ with one angular momentum turned on, and
explore solutions where we allow $W=0$.  In this case there are
two nontrivial Killing directions and one nontrivial cosine, like
the $D=4$ case.  A treatment of general $D$,
general $a_i$ will not be given here.

\subsection{Black Holes, S-branes:  Odd dimensions}

In odd dimensions $D=2n+1$, there are $n$ angular momentum
parameters $a_i$, $i=1,\ldots,n$ for a spinning black hole,
and we want to turn off just one of them, $a_n=0$.  This will
force a horizon/polar-origin/orbifold for
both the black hole and S-brane at $r^2=0$.

For $\lambda=1/l^2$ for $\Lambda=(D-1)\lambda>0$, the Kerr-dS solution of 
\cite{Gibbons:2004uw} is
\begin{eqnarray}\nonumber
ds^2&=&-W(1-\lambda r^2)(dx^D)^2+{U dr^2\over V-2M}+{2M\over U}
\Big(dx^D-\sum_{i=1}^{n-1}{a_i\mu_i^2 d\phi_i\over 1+\lambda a_i^2}\Big)^2\\
&&+\sum_{i=1}^n{r^2+a_i^2\over 1+\lambda a_i^2}\big(d\mu_i^2+\mu_i^2
(d\phi_i-\lambda a_i dx^D)^2\big)\nonumber\\
&&+{\lambda\over W(1-\lambda r^2)}\Big(\sum_{i=1}^n {(r^2+a_i^2)\mu_i d\mu_i
\over 1+\lambda a_i^2}\Big)^2.
\end{eqnarray}
where $V={1\over r^2}(1-\lambda r^2)\prod_{i=1}^n(r^2+a_i^2)$,
$W=\sum_{i=1}^n {\mu_i^2\over 1+\lambda a_i^2}$ and 
$U=\sum_{i=1}^n {\mu_i^2\over r^2+a_i^2}\prod_{j=1}^n (r^2+a_j^2)$.
The constraint is $\sum_{i=1}^n \mu_i^2=1$ and for reference,
$F=U/V={r^2\over 1-\lambda r^2}\sum_{i=1}^n {\mu_i^2\over r^2+a_i^2}$.
The thermodynamics of these black holes are discussed in \cite{Gibbons:2004ai}.

To get the S-brane, continue
$\mu_i\to i\mu_i$ for $i=1,\ldots,n-1$, $\phi_n\to i\phi_n$
and perform a flip $g_{\mu\nu}\to -g_{\mu\nu}$. The change in the
sign of the metric necessitates a change in the cosmological
constant --- our notation will be $\Lambda<0$ but $1/l^2=\lambda>0$.
So the S-Kerr-dS$_D$ solves Einstein's equations with $\Lambda<0$.
The constraint is now $\mu_n^2-\sum_{i=1}^{n-1}\mu_i^2=1$.
The solution is
\begin{eqnarray}\nonumber
ds^2&=&W(1-\lambda r^2)(dx^D)^2-{U dr^2\over V-2M}-{2M\over U}
\Big(dx^D+\sum_{i=1}^{n-1}{a_i\mu_i^2 d\phi_i\over 1+\lambda a_i^2}\Big)^2\\
&&+\sum_{i=1}^{n-1}{r^2+a_i^2\over 1+\lambda a_i^2}\big(d\mu_i^2+\mu_i^2
(d\phi_i-\lambda a_i dx^D)^2\big)-r^2(d\mu_n^2-\mu_n^2 d\phi_n^2)\nonumber\\
&&-{\lambda\over W(1-\lambda r^2)}\Big(r^2\mu_n d\mu_n-\sum_{i=1}^{n-1}{(r^2+a_i^2)\mu_i d\mu_i
\over 1+\lambda a_i^2}\Big)^2.
\end{eqnarray}
Now, $W$ and $U$ are given in terms of $\mu_i$ by
$$W=\mu_n^2-\sum_{i=1}^{n-1}{\mu_i^2\over 1+\lambda a_i^2}$$
and
$$U=\Big(\mu_n^2-\sum_{i=1}^{n-1}{\mu_i^2\over 1+a_i^2/r^2}\Big)
\prod_{j=1}^{n-1} (r^2+a_j^2).$$
Note that all elements in the solution that are functions of
$r$, $a_i$ and $\lambda$ have not changed and so it is clear that
the S-brane has the same parameter regions and
ensuing horizon structure as the black hole. However, the description 
of each $r$-interval as stationary or time-dependent is flipped. In particular, horizons are still
located at $V-2M=0$.  The determinant of
the $(n+1)\times (n+1)$ Killing submetric has
$\det g_{\alpha\beta}=r^2(V-2M)W\prod_{i=1}^n {\mu_i^2\over 1+\lambda a_i^2}$,
which has opposite sign from the black hole case \cite{Gibbons:2004uw}.

Consider the horizon function
$$r^2(V-2M)=r^2\Big((1-\lambda r^2)\prod_{i=1}^{n-1}(r^2+a_i^2)-2M\Big).$$
The geometry depends only on $r^2$, which we take to be the
independent variable, and allow to be negative.
In the ordinary parameter range $2M>\prod_{i=1}^{n-1}a_i^2=A^2$
then for sufficiently small $\lambda$, there are two positive
roots $0<r_1^2\leq r_2^2$. For $r^2>r_2^2$,
the solution is stationary as we expect for asymptotically AdS$_D$
space. With $r_1^2<r^2<r_2^2$ the solution is time-dependent and for 
$r_1^2=r_2^2$, there is an extremal horizon with an AdS$_2$ scaling limit.
For $0<r^2<r_1^2$ the solution is stationary.  At $r^2=0$, the solution
is not singular and the $\phi_n$-circle closes with
periodicity\footnote{The global structure here for $D$ odd is just like the
$\lambda=0$ case, which was first discovered by Jones and Wang \cite{cones}.}
$$\phi_n\simeq\phi_n+2\pi\sqrt{A^2/(A^2-2M)}.$$
Since $r^2\geq 0$, there is no $U=0$ singularity.

For the anomalous parameter
range $0<2M<A^2$ (described for $\lambda=0$ in \cite{cones} and in
great detail for $D=5$ black holes and S-branes), any $\lambda>0$ is
allowed and we have a root $r^2=r_2^2>0$, a $\phi_n$ Milne horizon
at $r^2=0$ and another root at $r^2=-r_1^2<0$.  Without loss of
generality we assume $0<a_1\leq a_2\leq\cdots\leq a_{n-1}$ and so 
$-a_1^2<-r_1^2<0$. The spacetime closes at $r^2=-r_1^2$ --- the
twisting and periodicity can be gotten by continuing the
angular momentum and surface gravity of
\cite{Gibbons:2004uw}.  In particular, just
put $r_{\rm horizon}\to ir_1$ ($r_{\rm horizon}\to -ir_1$ also gives
the right answer) and $\kappa\to \pm i\kappa$.  We have
$$\Omega^i={a_i(1+\lambda a_i^2)\over a_i^2-r_1^2},\qquad
\kappa=r_1(1+\lambda r_1^2)\sum_{i=1}^n {1\over a_i^2-r_1^2}+{1\over r_1}.$$
For $n=2$, $\lambda=0$, this reduces to $\Omega=a/2M$,
$\kappa={\sqrt{a^2-2M}\over 2M}$, which matches \cite{cones}.
Since $-r^2<a_1^2\leq a_2^2\leq\cdots$, there is no $U=0$ singularity.

The region $-r_1^2\leq r^2\leq 0$ for an anomalous range S-Kerr-dS
gives upon $\phi_n\to i\phi_n$ the new S-Kerr instanton of \cite{cones}.
The extremal case $2M=A^2$ is nonsingular at $r^2=r_1^2=0$ and has
a dS$_3$ scaling limit as described in \cite{cones}.

For the Kerr-dS black hole, $\lambda>0$ and $W=\sum_{i=1}^n {\mu_i^2\over 1+\lambda a_i^2}$ never zeroes.
For $\lambda$ a little negative, namely $-\min_i(a_i^{-2})<\lambda<0$,
$W$ is still positive.  This is the Kerr-AdS black hole which avoids $W=0$.
For $\lambda<-\min_i(a_i^{-2})$, there is a mixture of positive
and negative terms and we will find a $W=0$ coordinate singularity 
(moreover, a priori the spacetime has the wrong signature).

For S-Kerr-dS, we have $W=\mu_n^2-\sum_{i=1}^{n-1}{\mu_i^2\over 1+\lambda a_i^2}$.
For $\lambda>0$, from the constraint $\mu_n^2-\sum_{i=1}^{n-1}\mu_i^2=1$, this never zeroes;
we have a good S-Kerr-dS with $\Lambda<0$.
However, any $\lambda<0$ will give us a $W=0$ coordinate singularity.

\subsection{Black holes and S-branes:  Even dimensions}

The even-dimensional case $D=2n+2$ is quite different from the
odd-dimensional case.  Here we have $n$ rotation parameters and
we want to leave them all on, so the black hole has an equatorial
`ring' singularity
and the S-brane is nonsingular at $r=0$.\footnote{The global structure of this solution
for the $\lambda=0$ case was worked out by L\"u and V\'azquez-Poritz \cite{Lu:2004ye}.
The $D=4$ S-Kerr-dS with $\Lambda<0$ is in \cite{klemm}.}
The black hole solution is
\begin{eqnarray*}
ds^2&=&-W(1-\lambda r^2)(dx^D)^2+{Udr^2\over V-2M}+{2M\over U}\Big(dx^D-\sum_{i=1}^n
{a_i\mu_i^2d\phi_i\over 1+\lambda a_i^2}\Big)^2\\
&&+d\mu_0^2+\sum_{i=1}^n {r^2+a_i^2\over 1+\lambda a_i^2}d\mu_i^2
+\sum_{i=1}^n {r^2+a_i^2\over 1+\lambda a_i^2}\mu_i^2(d\phi_i-\lambda a_i dx^D)^2\\
&&+{\lambda\over W(1-\lambda r^2)}\Big(r^2\mu_0d\mu_0+\sum_{i=1}^n{(r^2+a_i^2)
\mu_i d\mu_i\over 1+\lambda a_i^2}\Big)^2.
\end{eqnarray*}
The constraint is $\mu_0^2+\sum_{i=1}^{n}\mu_i^2=1$, where $-1\leq\mu_0\leq 1$
has no rotation parameter and $0\leq\mu_i\leq 1$ for $i=1,\ldots,n$ has
rotation parameter $\phi_i$.  We have $V={1\over r}(1-\lambda r^2)\prod_{i=1}^n
(r^2+a_i^2)$, $W=\mu_0^2+\sum_{i=1}^n {\mu_i^2\over 1+\lambda a_i^2}$,
$U={1\over r}\Big(\mu_0^2+\sum_{i=1}^n{\mu_i^2\over 1+a_i^2/r^2}\Big)
\prod_{i=1}^n(r^2+a_i^2)$, and for reference
$F=U/V={r^2\over 1-\lambda r^2}\big(\mu_0^2+\sum_{i=1}^n{\mu_i^2\over r^2+a_i^2}\big)$.

For the even-dimensional case, the solution depends
properly on $r$, not $r^2$.  It has the
symmetry $r\to -r$, $M\to -M$.  For the right range of
parameters, the horizon function $rV(r)-2Mr$ has four roots,
$r_1<0<r_2\leq r_3\leq r_4$; an extremal black hole occurs for
$r_2=r_3$.  Since all $a_i$ are turned on,
at $r=0$ there is only a $U=0$ $S^{D-3}$ `ring' singularity at $\mu_0=0$ --- 
hence we can go to negative $r$.

The continuation to S-brane is $\mu_i\to \mu_i$, $i=1,\ldots,n$,
and $g_{\mu\nu}\to -g_{\mu\nu}$.  We then have $\Lambda$ with sign opposite
to $\lambda$.
\begin{eqnarray*}
ds^2&=&W(1-\lambda r^2)(dx^D)^2-{Udr^2\over V-2M}-{2M\over U}\Big(dx^D+\sum_{i=1}^n
{a_i\mu_i^2d\phi_i\over 1+\lambda a_i^2}\Big)^2\\
&&-d\mu_0^2+\sum_{i=1}^n {r^2+a_i^2\over 1+\lambda a_i^2}d\mu_i^2
+\sum_{i=1}^n {r^2+a_i^2\over 1+\lambda a_i^2}\mu_i^2(d\phi_i-\lambda a_i dx^D)^2\\
&&-{\lambda\over W(1-\lambda r^2)}\Big(r^2\mu_0d\mu_0-\sum_{i=1}^n{(r^2+a_i^2)
\mu_i d\mu_i\over 1+\lambda a_i^2}\Big)^2
\end{eqnarray*}
The quantity $V$ is as for the black hole, $W=\mu_0^2-\sum_{i=1}^n
{\mu_i^2\over 1+\lambda a_i^2}$ and $U={1\over r}\Big(\mu_0^2
-\sum_{i=1}^n{\mu_i^2\over 1+a_i^2/r^2}\Big)\prod_{i=1}^n(r^2+a_i^2)$.
The constraint is now $\mu_0^2-\sum_{i=1}^n\mu_i^2=1$ and this 
implies $\mu_0\geq 1$.  Since there is no ring singularity, 
we may follow $r\to -\infty$ and so the solution is nonsingular.

S-Kerr-dS with $\lambda>0$ avoids $W=0$.  S-Kerr-AdS with $\lambda<0$
hits a $W=0$ coordinate singularity.

The $r_2=r_3$ extremal case of the S-brane is nonsingular at the extremal horizon
and has a dS$_2$ scaling limit \cite{Lu:2004ye}.  The $r_3=r_4$ extremal
case has an AdS$_2$ scaling limit.

\subsection{Asymptotics}

Take say the odd $D$ case.  Sending $r\to\infty$ for
the black hole, we get a CBG
\begin{eqnarray*}
ds^2&\propto&\lambda W(dx^D)^2+\sum_{i=1}^n{1\over 1+\lambda a_i^2}
\big(d\mu_i^2+\mu_i^2(d\phi_i-\lambda a_i dx^D)^2\big)\\
&&-{1\over W}\big(\sum_{i=1}^n {\mu_i d\mu_i\over 1+\lambda a_i^2}\big)^2.
\end{eqnarray*}
This appears to be spinning, but the spinning is a pure coordinate effect.
If we let $\phi_i=\tilde{\phi_i}+\lambda a_i x^D$, we get
$$ds^2\propto\lambda W(dx^D)^2+\sum_{i=1}^n{1\over 1+\lambda a_i^2}
\big(d\mu_i^2+\mu_i^2d\tilde{\phi_i}^2\big)
-{1\over W}\big(\sum_{i=1}^n {\mu_i d\mu_i\over 1+\lambda a_i^2}\big)^2.$$
This is the same CBG as we get from the $M=0$ case, which was
identically (A)dS$_D$.  In fact these are just the `spheroidal coordinates'
of \cite{Gibbons:2004uw}.  This boundary is conformal
to ${\bf R}_{\rm space}\times S^{D-2}$ for $\lambda>0$ and
${\bf R}_{\rm time}\times S^{D-2}$ for $\lambda<0$.

The S-brane we know has no $W=0$ locus for $\lambda>0$ ($\Lambda<0$)
and has CBG
\begin{eqnarray*}
ds^2&\propto&\lambda W(dx^D)^2+\sum_{i=1}^{n-1}{1\over 1+\lambda a_i^2}
\big(d\mu_i^2+\mu_i^2(d\phi_i-\lambda a_i dx^D)^2\big)\\
&&-d\mu_n^2+\mu_n^2d\phi_n^2+{1\over W}\big(\mu_nd\mu_n-\sum_{i=1}^{n-1}{\mu_i d\mu_i\over 1+\lambda a_i^2}\big)^2.
\end{eqnarray*}
Again sending $r\to\infty$ has dropped out the $M$ parameter, so this
CBG should be conformal to ${\bf R}_{\rm time}\times{\bf H}_{D-2}$.

\subsection{The $\mu_0$-negative S-Kerr-dS for even dimensions}

There is another S-brane obtainable from Kerr-dS$_D$ for $D$ even.  This is
the analog of the 4d `Kerr $\pi/2$-bubble on its side' of \cite{cones}.
A sphere in even $D$ is written as
$$d\Omega_{D-2}^2=d\mu_0^2+d\mu_1^2+\cdots+d\mu_n^2+\mu_1^2d\phi_1^2
+\cdots+\mu_n^2 d\phi_n^2,$$
where the constraint is $\mu_0^2+\sum_{i=1}^{n}\mu_i^2=1$.
Send $\mu_0\to i\mu_0$ and $\mu_i\to i\mu_i$ $i=2,\ldots,n$,
$\phi_1\to i\phi_1$, and flip the sign of the metric.  We then get
$$d{\bf H}_{D-2}^2=d\mu_0^2-d\mu_1^2+\cdots+d\mu_n^2+\mu_1^2d\phi_1^2
+\cdots+\mu_n^2 d\phi_n^2,$$
where $-\mu_0^2+\mu_1^2-\mu_2^2-\cdots-\mu_n^2=1$.  In the Kerr case
along with $\phi_1\to i\phi_1$ we must do $a_1\to ia_1$.  We
call the resulting solution the $\mu_1$-positive S-Kerr-dS, or a
$\mu_0$-negative S-Kerr-dS. This emphasizes that in the ${\bf H}_{D-2}$
constraint, it is not $\mu_0$ but rather a $\mu_i$ that has a rotation
angle, that has the plus sign.  The full Kerr S-brane is
\begin{eqnarray*}
ds^2&=&W(1-\lambda r^2)(dx^D)^2-{Udr^2\over V-2M}-{2M\over U}\Big(dx^D
+{a_1\mu_1^2 d\phi_1\over 1-\lambda a_1^2}\\
&&+\sum_{i=2}^n{a_i\mu_i^2 d\phi_i\over 1+\lambda a_i^2}\Big)^2
+d\mu_0^2-{r^2-a_1^2\over 1-\lambda a_1^2}d\mu_1^2+\sum_{i=2}^n{r^2+a_i^2\over
1+\lambda a_i^2}d\mu_i^2\\
&&+{r^2-a_1^2\over 1-\lambda a_1^2}\mu_1^2(d\phi_1-\lambda a_1 dx^D)^2
+\sum_{i=2}^n {r^2+a_1^2\over 1+\lambda a_i^2}\mu_i^2(d\phi_i-\lambda a_i dx^D)^2\\
&&-{\lambda\over W(1-\lambda r^2)}\Big(-r^2\mu_0 d\mu_0+{r^2-a_1^2\over 1-\lambda a_1^2}
\mu_1 d\mu_1-\sum_{i=2}^n {r^2+a_i^2\over 1+\lambda a_i^2}\mu_i d\mu_i\Big)^2.
\end{eqnarray*}
Here, $W=-\mu_0^2+{\mu_1^2\over 1-\lambda a_1^2}-\sum_{i=2}^n{\mu_i^2
\over 1+\lambda a_i^2}$.  For $0<\lambda<1/a_1^2$, this does not zero, and
we never go to imaginary $\mu_1$.  There is no $U=0$ singularity at $r=a_1$,
because $\mu_1\geq 1$.  Horizons are given by $r_1\leq r_2<0<r_3\leq r_4$.
This solution is also important
for constructing even-dimensional anti-bubbles.

\subsection{Spinning $\Lambda$ bubble or anti-bubble solutions}

It was noticed in \cite{Balasubramanian:2002am} that Kerr-AdS bubbles
are `problematic' from the $W=0$ singularity, even in the simple 4d case.
We now check that $W=0$ always occurs in Kerr-AdS bubbles and find ways
to avoid $W=0$ for anti-bubbles constructed from Kerr-dS.

Bubbles with large-$r$ asymptotia can only come from $\lambda<0$ ($\Lambda<0$) black holes, where the large-$r$ region is stationary.  Then in even $D=2n+2$,
$$W=\mu_0^2+\sum_{i=1}^n{\mu_i^2\over 1+\lambda a_i^2}.$$
One way to get a bubble is to send $\mu_0\to i\mu_0$, $x^D\to ix^D$, and
$a_i\to ia_i$ $i=1,\ldots,n$.  This is the analog of the Kerr $\pi/2$-bubble
in 4d gotten from $\theta\to\pi/2+i\theta$, which has no Killing horizons or
CTCs.  Then
$$W=-\mu_0^2+\sum_{i=1}^n {\mu_i^2\over 1-\lambda a_i^2},\qquad{\rm where}
\qquad -\mu_0^2+\sum_{i=1}^n\mu_i^2=1.$$
We see that the terms with $+$ signs in $W$ are divided to make
them smaller; the $-\mu_0^2$ can dominate and make $W=0$.

Another type of bubble\footnote{These two types of bubbles are not the
same as the two solutions presented at the beginning of \cite{Ghezelbash:2002kf},
in the context of one angular momentum on.  There,
the first is a bubble with $W=0$, and the second is an anti-bubble with
$W=0$; its dS$_{D-5}$ is part of an AdS$_{D-4}$ which is part of a
perturbed AdS$_{D-2}$.  The construction
of an anti-bubble by continuing from hyperbolic space suggests
those authors also considered the S-brane.} is gotten by picking one angle from $\phi_1,\ldots,\phi_n$,
for example $\phi_1$.  Take $x^D\to ix^D$, $\phi_1\to i\phi_1$,
$a_i\to ia_i$ $i=2,\ldots,n$.  This is the analog of the ${\cal K}_+$ bubble
of \cite{Jones:2004pz,cones} in 4d obtained from double Killing continuation,
which has Killing horizons and CTCs.  In our present case, we can
continue past $\mu_1=0$ to $\mu_1\to i\mu_1$ and get
$$W=\mu_0^2-{\mu_1^2\over 1+\lambda a_1^2}+\sum_{i=2}^n {\mu_i^2\over
1-\lambda a_i^2},\qquad{\rm where}\qquad
\mu_0^2-\mu_1^2+\sum_{i=1}^n\mu_i^2=1.$$
Even if $a_1=0$, if some $a_i$ is turned on, $W=0$ still occurs.

For odd $D$, there is no $\mu_0$.  One can check that in any case except
no angular momentum, $W=0$ occurs.\footnote{In \cite{Balasubramanian:2002am},
(31)-(34) should be corrected to have $\Delta_\tau=1-{\alpha^2\over l^2}\sinh^2\tau
-{\beta^2\over l^2}\cosh^2\tau$, so $W\propto\Delta_\tau=0$ also occurs.}

For anti-bubbles, the situation is better --- we find 
Kerr-dS$_D$ anti-bubbles that avoid $W=0$ for all $D\geq 4$, and an extra one
in $D=4$.  The idea is to make the term with the $+$ coefficient in
$W$ to have a denominator smaller than the denominator of all $-$ terms.
Recall that to take an anti-bubble we start with Kerr-dS$_D$ with $\lambda>0$.
Take $D$ even and first go to the usual S-brane with $\Lambda<0$:
$$W=\mu_0^2-\sum_{i=1}^n{\mu_i^2\over 1+\lambda a_i^2},\qquad{\rm where}
\qquad \mu_0^2-\sum_{i=1}^n \mu_i^2=1.$$
Then pick say the angle $\phi_1$.  Send $x^D\to ix^D$, $\phi_1\to i\phi_1$,
$a_i\to ia_i$ $i=2,\ldots,n$.  We have
$$W=\mu_0^2-{\mu_1^2\over 1+\lambda a_1^2}-\sum_{i=2}^n{\mu_i^2\over 1-\lambda a_i^2},\qquad{\rm where}
\qquad \mu_0^2-\sum_{i=1}^n \mu_i^2=1.$$
This hits $W=0$ unless we turn off $a_i$ (with $i=2,\ldots,n$), but
in that case going through the horizon to $\mu_1\to i\mu_1$,
$$W=\mu_0^2+{\mu_1^2\over 1+\lambda a_1^2}-\sum_{i=2}^n \mu_i^2.$$
This still hits zero unless $D=4$ where $i=2,\ldots,n$ don't exist.
So the $D=4$ case with $a$ turned on, works.  This solution can be
easily obtained
by card diagram methods using Fig.~\ref{RNdS4fig} by performing
a $\gamma$-flip on the $r\geq r_4$, $0\leq\theta\leq\pi$ region (and
extending to $\theta\to i\theta$ or $\theta\to\pi+i\theta$).

In even $D$, we can also find a whole family of anti-bubbles from
$\mu_0$-negative S-branes.  Picking $\mu_1$ to have the $+$ sign, this S-brane
has
$$W=-\mu_0^2+{\mu_1^2\over 1-\lambda a_1^2}-\sum_{i=2}^n{\mu_i^2\over 1+\lambda a_i^2},\qquad{\rm where}
\qquad -\mu_0^2+\mu_1^2-\sum_{i=2}^n \mu_i^2=1.$$
Then double Killing $x^D\to ix^D$, $\phi_1\to i\phi_1$, $a_i\to ia_i$
(with $i=2,\ldots,n$), we have
$$W=-\mu_0^2+{\mu_1^2\over 1-\lambda a_1^2}-\sum_{i=2}^n{\mu_i^2\over 1-\lambda a_i^2},\qquad{\rm where}
\qquad -\mu_0^2+\mu_1^2-\sum_{i=2}^n \mu_i^2=1.$$
For $0<1-\lambda a_1^2<1-\lambda a_i^2$ $i=2,\ldots,n$, we avoid $W=0$.
Note that we never get to a $\mu_1=0$ $\phi_1$-horizon, so the above distribution
of hyperbolic pieces is global.  So for general even $D$ we have this Kerr-dS
anti-bubble.  We stress that the $D=4$ solution of this
is different from the one gotten from the ordinary S-brane.  This present $D=4$ solution
can be obtained from the black hole by $\theta\to \pi/2+i\theta$, performing
a $\gamma$-flip to make the non-Killing directions $++$, and then
$\phi\to i\phi$, $a\to ia$.  To avoid the $U=0$ `ring' singularity, taking
$a_1>a_2\geq\cdots$, we want the largest horizon root (the anti-bubble)
to occur at $r_4>a_2$.  This can always be arranged for large enough $l$.

In odd dimension $D=2n+1$, the S-brane with $a_n$ turned back on has
$$W={\mu_n^2\over 1-\lambda a_n^2}-\sum_{i=1}^{n-1}{\mu_i^2\over 1+\lambda a_i^2},\qquad{\rm where}\qquad
\mu_n^2-\sum_{i=1}^{n-1}\mu_i^2=1.$$
To get a good anti-bubble, we send $x^D\to ix^D$, $\phi_n\to i\phi_n$, $a_i\to ia_i$ $i=1,\ldots,n-1$, hence
$$W={\mu_n^2\over 1-\lambda a_n^2}-
\sum_{i=1}^{n-1}{\mu_i^2\over 1-\lambda a_i^2}.$$
For $0<1-\lambda a_n^2<1-\lambda a_i^2$, we avoid $W=0$.  Again,
we never reach a $\mu_n=0$ $\phi_n$-horizon so the above characterization
is global.  The $D=5$ solution can be obtained by $\gamma$-flipping
the $r\geq r_4$, $0\leq\theta\leq\pi/2$ of the black hole, going
to $\mu_1\to i\mu_1$ where $\mu_1=\sin\theta$, then continuing
$\phi_i\to i\phi_i$, $a_i\to ia_i$ (with $i=1,2$).  Taking $a_n^2>a_1^2\geq\cdots$,
the largest root (anti-bubble) is at $r_4^2>a_1^2$ for large $l$, so
the solution is nonsingular.

\section{Kerr-(A)dS$_D$: One $a_i$ on, and allow $W=0$}
\label{wzero}

When a $W=0$ coordinate singularity occurs, an extra horizon-like
locus will be present.  We will just look at the case of one
angular momentum on, where the Kerr-(A)dS$_D$ solution simplifies
considerably \cite{Hawking:1998kw}.
For bubbles, we find that in $D\geq 4$ there is
one family of nonsingular solutions and in $D=4$ there is an
additional solution.  For S-Kerr-AdS, $D\geq 4$ we find
one family.  There are no other generically nonsingular solutions.

\subsection{Bubbles}

Let's examine bubbles; take $D=4$ first.  The Kerr-AdS$_4$ black hole
solution \cite{Carter:1968ks} is
\begin{eqnarray}\nonumber
ds^2&=&\rho^2\big({dr^2\over\Delta}+{d\theta^2\over 1-(a^2/l^2)\cos^2\theta}\big)-{\Delta\over \rho^2}\big(dx^4-{a\sin^2\theta\over 1-a^2/l^2}d\phi\big)^2\\
&&+{\sin^2\theta(1-(a^2/l^2)\cos^2\theta)\over \rho^2}\big(adx^4-{r^2+a^2\over
1-a^2/l^2}d\phi\big)^2,\label{kerrads4}
\end{eqnarray}
where $\Delta=(r^2+a^2)(1+r^2/l^2)-2Mr$, $\rho^2=r^2+a^2\cos^2\theta$,
and $0\leq\theta\leq \pi$.

The $\pi/2$-bubble is gotten from $\theta\to\pi/2+i\theta$, $a\to ia$, $x^4\to ix^4$.
Then
\begin{eqnarray}\nonumber
ds^2&=&\rho^2\big({dr^2\over\Delta}-{d\theta^2\over 1-(a^2/l^2)\sinh^2\theta}\big)+{\Delta\over \rho^2}\big(dx^4-{a\cosh^2\theta\over 1+a^2/l^2}d\phi\big)^2\\
&&+{\cosh^2\theta(1-(a^2/l^2)\sinh^2\theta)\over \rho^2}\big(adx^4+{r^2-a^2\over
1+a^2/l^2}d\phi\big)^2,
\end{eqnarray}
where $\Delta=(r^2-a^2)(1+r^2/l^2)-2Mr$ has roots $r_1<0<r_2$, and $\rho^2=r^2+a^2\sinh^2\theta$.
At $r=r_2$, $\Delta=0$ and the differential displacement
$adx^4+{r_2^2-a^2\over 1+a^2/l^2}d\phi=0$ is
null.  So let $\tilde\phi=\phi-\Omega x^4$, $\tilde x^4=x^4$, $\Omega=-a(1+a^2/l^2)
/(r_2^2-a^2)$.  Then $d\tilde\phi=0$ is null so the vector
$(\partial/\partial\tilde x^4)_{\tilde\phi}$ is null at $r=r_2$.
We can compactify $\tilde x^4\simeq \tilde x^4+\beta$ for some
periodicity to make $r=r_2$ the origin of polar coordinates.  We
must leave $\tilde\phi$ noncompact to get a horizon at $W=0$ (and
make no reference to the previous $x^4\phi$ Killing
coordinates).\footnote{The differentials $dx^4$, $d\phi$ are still well defined
and it is still acceptable to write the metric in terms of them.}
The metric is now
\begin{eqnarray}\nonumber
ds^2&=&\rho^2\big({dr^2\over\Delta}-{d\theta^2\over 1-(a^2/l^2)\sinh^2\theta}\big)+{\Delta\over \rho^2}\big(dx^4-{a\cosh^2\theta\over 1+a^2/l^2}(d\tilde\phi+\Omega d\tilde x^4)\big)^2\\
&+&{\cosh^2\theta(1-(a^2/l^2)\sinh^2\theta)\over \rho^2}\big(ad\tilde x^4
+{r^2-a^2\over 1+a^2/l^2}(d\tilde\phi+\Omega d\tilde x^4)\big)^2.
\end{eqnarray}
Setting $\sinh\theta_0=l/|a|$, 
this metric has expected bubble properties for $-\theta_0<\theta<\theta_0$.
Following \cite{Balasubramanian:2002am}, set $\sinh\theta=l/a-\epsilon^2$;
then for small real $\epsilon$, we have
\begin{eqnarray}\nonumber
ds^2&\approx&(r^2+l^2){dr^2\over\Delta}-{2la(l^2+r^2)\over l^2+a^2}d\epsilon^2
+{\Delta\over r^2+l^2}\big(d\tilde x^4-{l^2\over a}
(d\tilde\phi+\Omega d\tilde x^4)\big)^2\\
&&+\epsilon^2 {2(l^2+a^2)\over la (l^2+r^2)}\big(ad\tilde x^4
+{r^2-a^2\over 1+a^2/l^2}(d\tilde\phi+\Omega d\tilde x^4)\big)^2.
\end{eqnarray}
At $\epsilon=0$, $d\tilde x^4-{l^2\over a}(d\tilde\phi+\Omega d\tilde x^4)=0$
is null. That is a regular spinning horizon --- the angular 
velocity of a regular spinning horizon must be constant \cite{carterleshouches,Gibbons:2004uw}.
On this side of the horizon the Killing
submetric has signature $++$ and on the other side it will have $+-$.
We let $\tilde{\tilde x}^4=\tilde x^4-\mho\tilde\phi$, $\tilde{\tilde\phi}=\tilde\phi$,
where $\mho=\Omega^{-1}(1-a^2/l^2)$.  Then $d\tilde{\tilde x}^4=0$ is
null, or the vector $(\partial/\partial\tilde{\tilde\phi})_{\tilde{\tilde x}^4}$
gives us the Milne trajectories.

\begin{figure}[htb]
\begin{center}
\epsfxsize=1.5in\leavevmode\epsfbox{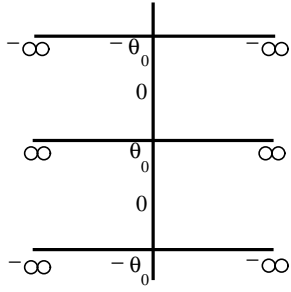}
\caption{A skeleton diagram for the $\theta$ coordinate
of the Kerr-AdS$_4$ $\pi/2$-bubble. Horizontal segments are where
$\theta$ is spacelike and vertical where $\theta$ is timelike.
Here, four-segment intersections are spinning $\tilde x^4 \tilde\phi$
horizons.}
\label{kbubble1fig}
\end{center}
\end{figure}

This horizon is then just like a Kerr horizon except the role of the two
Killing metric squares is reversed and $r$ is replaced by $\theta$.  If
we repress the Killing directions and $r\geq r_2$, we arrive at a spacetime
skeleton diagram for just the $\theta$ coordinate --- see Fig.~\ref{kbubble1fig}.  The
vertical legs have $\theta$ timelike and the horizontal legs have
$\theta$ spacelike.  The spacetime
is periodic in time; each dS$_{D-2}$-type region gives way to horizons beyond which
are stationary regions.  The $r\to\infty$ limit gives a CBG which can
be represented by the same skeleton diagram --- the metric is
$$ds^2\propto {1\over l^2}\big(d\tilde x^4-{a\cosh^2\theta\over 1+a^2/l^2}(d\tilde\phi
+\Omega d\tilde x^4)\big)^2+{\cosh^2\theta\over 1+a^2/l^2}(d\tilde\phi+\Omega
d\tilde x^4)^2-{d\theta^2\over 1-{a^2\over l^2}\sinh^2\theta}.$$

On the other hand, we can form the double Killing bubble from (\ref{kerrads4}).
We take $x^4\to ix^4$ and $\phi\to i\phi$ and then the metric becomes
\begin{eqnarray}\nonumber
ds^2&=&\rho^2\big({dr^2\over\Delta}+{d\theta^2\over 1-{a^2\over l^2}\cos^2\theta}\big)
+{\Delta\over\rho^2}\big(dx^4-{a\sin^2\theta\over 1-a^2/l^2}d\phi\big)^2\\
&&-{\sin^2\theta (1-(a^2/l^2)\cos^2\theta)\over\rho^2}\big(adx^4
-{r^2+a^2\over 1-a^2/l^2}d\phi\big)^2,\label{kkillbub}
\end{eqnarray}
where $\Delta=(r^2+a^2)(1+r^2/l^2)-2Mr$, $\rho^2=r^2+a^2\cos^2\theta$.

We must twist like
before: $\tilde\phi=\phi-\Omega x^4$, $\tilde x^4=x^4$, $\Omega=a(1-a^2/l^2)/(r_2^2+a^2)$, $\tilde x^4\simeq\tilde
x^4+\beta$ for some $\beta$, $\tilde\phi$ noncompact.  So replace $d\phi\to d\tilde\phi+\Omega d\tilde x^4$ and $dx^4\to d\tilde x^4$ in (\ref{kkillbub}).

\begin{figure}[htb]
\begin{center}
\epsfxsize=4in\leavevmode\epsfbox{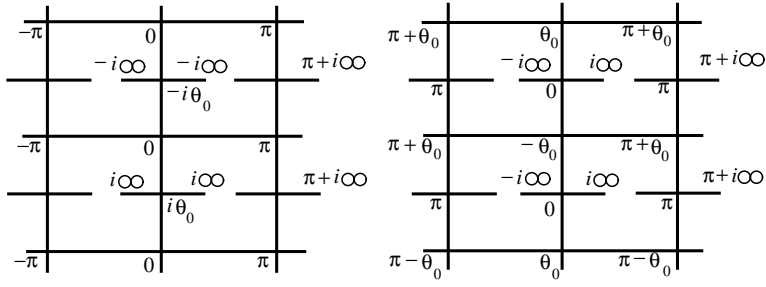}
\caption{Skeleton diagram for the $\theta$ coordinate
of the Kerr-AdS$_4$ double Killing bubble.  Horizontal segments are where
$\theta$ is spacelike and vertical where $\theta$ is timelike.
(a) The figure on the left is the case where $l^2>a^2$.  (b) The
figure on the right is the case where $l^2<a^2$.}
\label{kbubble2fig}
\end{center}
\end{figure}

Take the case $l^2>a^2$.
Then $0\leq\theta\leq\pi$
is fine, and we can continue $\theta\to i\theta$:
\begin{eqnarray}\nonumber
ds^2&=&\rho^2\big({dr^2\over\Delta}-{d\theta^2\over 1-{a^2\over l^2}\cosh^2\theta}\big)
+{\Delta\over\rho^2}\big(d\tilde x^4+{a\sinh^2\theta\over 1-a^2/l^2}(d\tilde\phi
+\Omega d\tilde x^4)\big)^2\\
&+&{\sinh^2\theta (1-(a^2/l^2)\cosh^2\theta)\over\rho^2}\big(ad\tilde x^4
-{r^2-a^2\over 1-a^2/l^2}(d\tilde \phi+\Omega d\tilde x^4)\big)^2,
\end{eqnarray}
where $\Delta=(r^2+a^2)(1+r^2/l^2)-2Mr$, $\rho^2=r^2+a^2\cosh^2\theta$.
With $\cosh^2\theta_0=l^2/a^2$, $\theta=\theta_0$ is a horizon as before,
and beyond it we have stationary regions of $\theta>\theta_0$.  The
skeleton diagram is shown in Fig.~\ref{kbubble2fig}(a).  It is canonical
to identify every other $0\leq\theta\leq\pi$ horizontal segment.  If
on the other hand $a^2>l^2$, then with $\cos^2\theta=l^2/a^2$, we imagine
expanding out from $\theta=\pi/2$, we have
$\theta_0$ occurring before $\theta=0,\pi$ and the skeleton diagram
is shown in Fig.~\ref{kbubble2fig}(b).  These solutions are nonsingular.

\begin{figure}[htb]
\begin{center}
\epsfxsize=2in\leavevmode\epsfbox{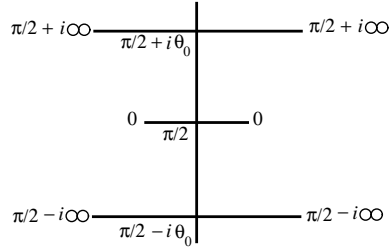}
\caption{A skeleton diagram for the $\theta$ coordinate
of the Kerr-AdS$_D$ $\pi/2$-bubble for $D\geq 5$, one turned on.
At $\theta=0$, a Killing direction closes the spacetime.  (For $D\geq 6$, there
is only one dS$_{D-4}$ leg and we have a $\vdash$ junction instead of $+$
junction at $\theta=\pi/2$.)
At $\theta=\pi/2$, dS$_{D-4}$ becomes null and becomes ${\bf H}_{D-4}$.
At $\theta=\pi/2\pm i\theta_0$, there is a 
spinning horizon orbifold --- this solution is singular.}
\label{kbubble3fig}
\end{center}
\end{figure}

For $D\geq 5$, we must add $r^2\cos^2\theta d\Omega_{D-4}^2$ to the metric as in (\ref{kerrads4}) \cite{Hawking:1998kw} and also set $\Delta=(r^2+a^2)(1+r^2/l^2)-2M/r^{D-5}$.
The motivation is
$$d\Omega_{D-2}^2=d\theta^2+\sin^2\theta d\phi^2+\cos^2\theta d\Omega_{D-4}^2,$$
where $0\leq\theta\leq\pi/2$.  To get the $\pi/2$-bubble, send
$\theta\to\pi/2+i\theta$, $x^D\to ix^D$, $a\to ia$, and $d\Omega_{D-4}^2\to
-d{\bf H}_{D-4}^2$.  Our motivating element becomes
$$d{\rm dS}_{D-2}^2=-d\theta^2+\cosh^2\theta d\phi^2+\sinh^2\theta d{\bf H}_{D-4}^2.$$
Alternatively, we could have done $d\Omega_{D-4}^2\to d{\rm dS}_{D-4}^2$, 
$x^D\to ix^D$, $a\to ia$, where our motivating element becomes
\begin{equation}\label{kdsbubbleD4}
d{\rm dS}_{D-2}^2=d\theta^2+\sin^2\theta d\phi^2+\cos^2\theta d{\rm dS}_{D-4}^2.
\end{equation}
These two representations of dS$_{D-2}$ are connected together
where either hyperbolic space or dS$_{D-4}$ degenerates to the null cone.
A skeleton diagram is drawn
in Fig.~\ref{kbubble3fig}.  This spacetime, however, is 
problematic --- we have a compactification at $\theta=0$ of (\ref{kdsbubbleD4}) and another compactification condition at $r=r_2$.
Generically, both Killing directions are compact, and
the horizon at $\theta=\pi/2\pm i\theta_0$ is thus an orbifold.

\begin{figure}[htb]
\begin{center}
\epsfxsize=4in\leavevmode\epsfbox{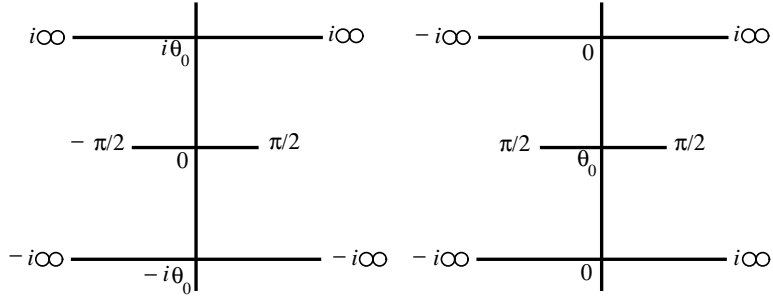}
\caption{Skeleton diagrams for the $\theta$ coordinate
of the Kerr-AdS$_D$ double-Killing bubble for $D\geq 5$, one turned on.
At $\theta=\pm\pi/2$, the $S^{D-4}$ closes the spacetime.
The four-leg junctions are all spinning horizons. (a) The
diagram on the left is for $l^2>a^2$.  (b) The diagram on the
right is for $l^2<a^2$.}
\label{kbubble4fig}
\end{center}
\end{figure}

The double-Killing bubble, gotten from $x^D\to ix^D$, $a\to ia$, is nonsingular.
There are two cases, $l^2>a^2$ and $a^2>l^2$.
The skeleton diagrams are different from the $D=4$ case and are shown in
Figs.~\ref{kbubble4fig}(a,b).

\subsection{S-branes and anti-bubbles}

The $\pi/2$-S-Kerr-AdS with one on can be motivated from the continuation
and sign flip
\begin{eqnarray*}
d\Omega_{D-2}^2&=&d\theta^2+\sin^2\theta d\phi^2+\cos^2\theta d\Omega_{D-4}^2\\
\rightarrow d{\bf H}_{D-2}^2&=&d\theta^2+\cosh^2\theta d\phi^2+\sinh^2\theta d\Omega_{D-4}^2.
\end{eqnarray*}
The full continuation of Kerr-AdS is $\theta\to\pi/2+i\theta$, $g_{\mu\nu}\to
-g_{\mu\nu}$, $\phi\to i\phi$, $a\to ia$.  The result has $\Lambda>0$ and is
\begin{eqnarray}
ds^2&=&\rho^2\big(-{dr^2\over\Delta} +{d\theta^2\over 1-(a^2/l^2)\sinh^2\theta}\big)
+{\Delta\over\rho^2}\big(dx^D+{a\cosh^2\theta\over 1+a^2/l^2}d\phi\big)^2\\
&+&{\cosh^2\theta\over\rho^2}(1-(a^2/l^2)\sinh^2\theta)\big(a dx^D
-{r^2-a^2\over 1+a^2/l^2}d\phi\big)^2+r^2\sinh^2\theta d\Omega_{D-4}^2,\nonumber
\end{eqnarray}
where $\rho^2=r^2+a^2\sinh^2\theta$ and $\Delta=(r^2-a^2)(1+r^2/l^2)-2M/r^{D-5}$.
This has $W=0$ at $\theta=\pm\theta_0$ where $\sinh\theta_0=l/|a|$; $\theta$ is spacelike and for $r>r_2$, the Killing directions are $++$, so this closes the spacetime.
The conditions at $\pm\theta_0$ are identical hence 
compatible and one Killing direction is noncompact to 
give a horizon at $r=r_2$.  This solution is nonsingular.

The $\pi/2$-S-Kerr-dS solution
has no $W=0$ and has already been discussed,
but we use $\pi/2$-S-Kerr-dS to construct anti-bubbles.
The $\pi/2$-S-Kerr-dS$_D$ solution with one on is
\begin{eqnarray}
ds^2&=&\rho^2\big(-{dr^2\over\Delta} +{d\theta^2\over 1+(a^2/l^2)\sinh^2\theta}\big)
+{\Delta\over\rho^2}\big(dx^D+{a\cosh^2\theta\over 1-a^2/l^2}d\phi\big)^2\\
&+&{\cosh^2\theta\over\rho^2}(1+(a^2/l^2)\sinh^2\theta)\big(a dx^D
-{r^2-a^2\over 1-a^2/l^2}d\phi\big)^2+r^2\sinh^2\theta d\Omega_{D-4}^2,\nonumber
\end{eqnarray}
where $\rho^2=r^2+a^2\sinh^2\theta$ and $\Delta=(r^2-a^2)(1-r^2/l^2)-2M/r^{D-5}$.

One anti-bubble is gotten from $x^D\to ix^D$, $\phi\to i\phi$, with
motivating element
$$d{\rm AdS}_{D-2}^2=d\theta^2-\cosh^2\theta d\phi^2+\sinh^2\theta d\Omega_{D-4}^2.$$
This solution does not have $W=0$ so it was already covered in the last section.

For $D\geq 5$, another anti-bubble is gotten from $x^D\to ix^D$, $a\to ia$, $d\Omega_{D-4}^2\to d{\rm dS}_{D-4}^2$.  Since $W\propto 1-(a^2/l^2)\sinh^2\theta$, the space closes
at $\theta=\pm\theta_0$.  But the space also closes at $r=r_2$ and these
two conditions are not compatible, orbifolding the horizon that occurs
at $\theta=\pm i\pi/2$.

We now investigate the usual S-Kerr-AdS and Kerr-dS anti-bubbles 
instead of the $\pi/2$-versions and find that generically they 
are all problematic, though there may be lower-dimensional parametric families or special cases that work.

The usual S-brane is gotten from the black hole by $g_{\mu\nu}\to -g_{\mu\nu}$,
$\theta\to i\theta$, $d\Omega_{D-4}^2\to -d{\bf H}_{D-4}^2$.  S-Kerr-AdS$_D$
with one on is
\begin{eqnarray}\label{skerrads}
ds^2&=&\rho^2\big(-{dr^2\over\Delta}+{d\theta^2\over 1-(a^2/l^2)\cosh^2\theta}\big)+{\Delta\over \rho^2}\big(dx^4+{a\sinh^2\theta\over 1-a^2/l^2}d\phi\big)^2\\
&+&{\sinh^2\theta(1-(a^2/l^2)\cosh^2\theta)\over \rho^2}\big(adx^4-{r^2+a^2\over
1-a^2/l^2}d\phi\big)^2+r^2\cosh^2\theta d{\bf H}_{D-4}^2,\nonumber
\end{eqnarray}
where $\rho^2=r^2+a^2\cosh^2\theta$ and $\Delta=(r^2+a^2)(1+r^2/l^2)-2M/r^{D-5}$.
Here $W=0$ occurs, and this solution is problematic.  Assuming
$l^2>a^2$ to get the right signature, the
$\theta=0$ and $\theta=\theta_0$ conditions
are incompatible, forcing $x^D$ to be compact and the $r=r_2$ horizon to
be an orbifold.

S-Kerr-dS$_D$ is
\begin{eqnarray}\label{skerrds}
ds^2&=&\rho^2\big(-{dr^2\over\Delta}+{d\theta^2\over 1+(a^2/l^2)\cosh^2\theta}\big)+{\Delta\over \rho^2}\big(dx^4+{a\sinh^2\theta\over 1+a^2/l^2}d\phi\big)^2\\
&+&{\sinh^2\theta(1+(a^2/l^2)\cosh^2\theta)\over \rho^2}\big(adx^4-{r^2+a^2\over
1+a^2/l^2}d\phi\big)^2+r^2\cosh^2\theta d{\bf H}_{D-4}^2,\nonumber
\end{eqnarray}
where $\rho^2=r^2+a^2\cosh^2\theta$ and $\Delta=(r^2+a^2)(1-r^2/l^2)-2M/r^{D-5}$.

The double Killing anti-bubble is gotten from (\ref{skerrds}) by
$x^D\to ix^D$, $\phi\to i\phi$.  The solution as written is then good
down to $\theta=0$ where we have a spinning Rindler horizon; then
move up to $\theta=\pm i\pi/2$ where ${\bf H}_{D-4}$ becomes 
dS$_{D-4}$ and then to $\theta=\pm i\pi/2\pm \theta_0$ with $\sinh\theta_0=l/|a|$,
where the space closes. The space closing here is generally 
incompatible with the
$r=r_2$ condition, making the $\theta=0$ horizon into an orbifold.
The exception is $D=4$ where there is no ${\bf H}_{D-4}$; this has
no $W=0$ and has already been discussed.

On the other hand, for $D\geq 5$,
making an anti-bubble from $x^D\to ix^D$, $a\to ia$, $d{\bf H}_{D-4}^2\to d{\rm AdS}_{D-4}^2$
gives $W\propto 1-(a^2/l^2)\cosh^2\theta$.  Assuming $l^2>a^2$, the spacetime closes at $\theta=0$
as well as $\theta=\theta_0$ and in general this is not compatible with the $r=r_2$ 
condition.  Also
there may be a `ring' singularity $\rho^2=0$, although it does not propagate to large $r$.

\section{Conclusions and Relation to Holography}
\label{sixthsec}

In this paper we presented a procedure to generate time-dependent 
(and other black and anti-bubble)
backgrounds starting from black holes solutions in (A)dS spacetime.
We hope that our unified treatment of S-branes, bubbles and anti-bubbles
with an emphasis on which solutions are possible, which are distinct, and
what is their global structure including horizons and singularities,
is useful to the reader.  Some solutions
in this paper are already known; several have been reexamined, reinterpreted or
renamed (the `anti-bubble') and several new solutions have been presented.

We have emphasized $D=4,5$ $r\theta$-diagrams and $\theta$-skeleton diagrams to
keep track of spacetime regions and for pure (A)dS$_D$ for $D=4,5$ we present
various card diagrams in Appendix A. Our analytic continuation has
been simple, involving only Killing directions and cosine directions.
For $D=4,5$ analytic continuation has been restated in terms of the
card diagram technique of the $\gamma$-flip.

We find six types of spacetimes with a characteristic expected
conformal boundary geometry. Black holes in AdS have $S^{D-2}\times 
{\bf R}_{\rm time}$, bubbles have dS$_{D-2}\times S^1$, anti-bubbles have AdS$_{D-2}\times S^1$ and
S-branes with $\Lambda<0$ have ${\bf H}_{D-2}\times {\bf R}_{\rm time}$. Black holes in dS have conformal boundary geometry $S^{D-2}\times {\bf R}_{\rm space}$ and S-branes with $\Lambda>0$ have ${\bf H}_{D-2}\times {\bf R}_{\rm space}$. Solutions from Kerr-(A)dS which have $W=0$ horizons, if they are good spacetimes, have a more complicated global structure for themselves and for their conformal boundaries.

Since many of the presented solutions are locally 
asymptotically (A)dS, it would be interesting to 
study them in the context of gauge/gravity dualities --- the holographic results concerning some of the new spacetimes are 
forthcoming. The main tool that we use is the counterterm method 
proposed by Balasubramanian and Kraus in \cite{Balasubramanian:1999re}.
That is, 
to regularize the boundary stress tensor and the gravity action 
by supplementing the quasilocal formalism \cite{Brown:1992br}
with counterterms depending of the {\it intrinsic} boundary geometry.
This way, the infrared divergencies of quantum gravity in the bulk 
are equivalent to ultraviolet divergences of dual theory living 
on the boundary.  This method was also generalized to locally 
asymptotically dS spacetimes \cite{Ghezelbash:2001vs, Klemm:2001ea}.
However, unlike  the AdS/CFT correspondence, the conjectured dS/CFT correspondence is far from being understood (see, e.g., 
\cite{Klemm:2004mb} for a nice review).

Recently, Ross and Titchener \cite{Ross:2004cb} used the counterterm 
method to show that the AdS/CFT may teach us how to choose the right 
vacuum for the strongly-coupled CFT living on a dS background. The 
main idea is to use the black hole-AdS-bubble solution 
as a laboratory for studying the description of vacuum ambiguities in AdS/CFT.  Also, Balasubramanian et al. 
\cite{Balasubramanian:2005bg} investigate the semiclassical decay 
of a class of orbifolds of AdS space via a bubble of 
nothing.

Using similar `holographic' reasoning\footnote{Other interesting 
examples of time-dependent AdS/CFT and dS/CFT correspondences can 
be found in \cite{Balasubramanian:2002am,Ghezelbash:2002kf,Astefanesei:2003gw,
Balasubramanian:2001nb, Cvetic:2003zy, Ghezelbash:2002xt}. It is 
also worth mentioning that, in a different context 
\cite{eu}, some unexpected results were obtained for asymptotically 
AdS Taub-NUT spacetimes.} to 
investigate some of the solutions presented in this paper, we hope 
to shed light on different aspects of the gauge/gravity correspondence for 
time-dependent backgrounds.

\acknowledgments
The authors would like to thank B. Julia, A. Maloney, E. Radu, 
A. Strominger, J. E. Wang and X. Yin for useful conversations. 
We also thank IAS, Princeton, where the work was initially discussed.
G.C.J. thanks the NSF for funding.

\appendix

\section{Generalized card diagrams for (A)dS$_4$, (A)dS$_5$}

Some of the solutions in this paper were found in analogy with
card diagram techniques \cite{Jones:2004pz} (see also
\cite{Weylpaper,EmparanWK,EmparanBB,Jones:2004rg,Harmark:2004rm}).
Furthermore the asymptotia of these solutions can be understood from
the card diagram perspective.
It is thus appropriate
to give a small application of card diagrams to (anti-)de Sitter
space in dimensions 4 and 5, where they have the requisite 2 and 3
commuting Killing fields.  The Weyl technique for Einstein's dynamical
equation fails with a nonzero $\Lambda$.  Nonetheless these spacetimes
still have satisfying card diagrams.  Here, we will not give a theory
of generalized card diagrams, but rather just some examples which
we can obtain by formal analogy to the 4d Reissner-Nordstr\o m black
hole.  More conventional
Penrose diagrams for (A)dS may be found in \cite{HawkingEllis,Spradlin:2001pw}.

The massless RN black hole of imaginary charge (to make it subextremal)
has line element $\propto{dr^2\over r^2-Q^2}+d\theta^2$.
Once the non-Killing directions are of this form, we can immediately go
to spherical prolate coordinates via $r=Q\cosh\zeta$; then $ds^2\propto
d\zeta^2+d\theta^2$; and then to card diagram coordinates via $\rho=Q\sinh\zeta\sin\theta$, $z=Q\cosh\zeta\cos\theta$.

\begin{figure}[htb]
\begin{center}
\epsfxsize=3.5in\leavevmode\epsfbox{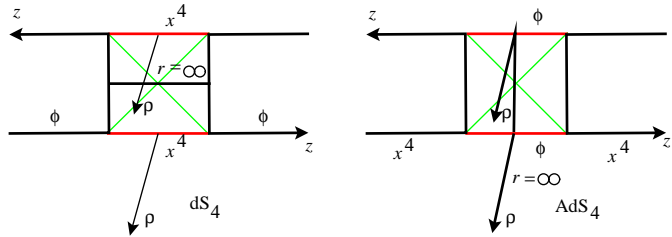}
\caption{(a) On the left, we have dS$_4$ fibered by $S^2$.  Turning the
vertical square card on its side, we get (b) the diagram on the right;
it is AdS$_4$ fibered by patched AdS$_2$.  This is the same fibering
as the RNdS$_4$ anti-bubble.}
\label{dScard1fig}
\end{center}
\end{figure}

De Sitter 4-space has ${dr^2\over 1-r^2/l^2}+d\theta^2$.  Set
$u=1/r$.  Then we get $\propto {du^2\over u^2-l^2}+d\theta^2$ and
can proceed as above.  The result is an elliptic card diagram
with a rod horizon $-1/l<z<1/l$, and the vertical square card
above it is bisected halfway up at $u=0$ (see Fig.~\ref{dScard1fig}(a)).
Please note that for simplicity we have only drawn two cards at
each 4-card horizon; see \cite{Jones:2004pz}.

Turning the vertical card on its side via the $\gamma$-flip, we achieve
AdS$_4$ in a coordinate system similar to the RNdS$_4$ anti-bubble
solution (see Fig.~\ref{dScard1fig}(b)).  An infinite stack of cards give periodic time.
The RNdS$_4$ anti-bubble asymptotes to all the $r=\infty$ asymptotia drawn
here.  Fig.~\ref{dScard1fig}(b) can be double Killing continued to give
a card diagram suitable for understanding S-RNdS$_4$; this will
be skipped for brevity.

\begin{figure}[htb]
\begin{center}
\epsfxsize=2.5in\leavevmode\epsfbox{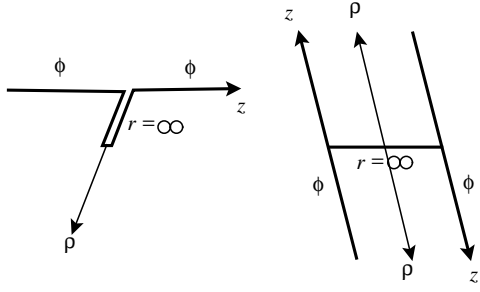}
\caption{AdS$_4$ fibered by $S^2$.  Shown as a doubly covered half-plane
and then in a conformally fixed picture.}
\label{AdScard1fig}
\end{center}
\end{figure}

To get AdS$_4$ fibered with spheres, we start with ${dr^2\over 1+r^2/l^2}+d\theta^2$.
Let $u=1/r$ as for de Sitter, and we get $\propto {du^2\over u^2+l^2}+d\theta^2$.
Now, the solution is superextremal and on a branched horizontal card.
To go to spherical prolate coordinates, let $u=l\sinh\zeta$.  The
resulting card diagram is shown in Fig.~\ref{AdScard1fig}.  This
card diagram can be double Killing continued to give AdS$_4$ fibered by dS$_2$,
like the RNAdS$_4$ bubble; this will be skipped for brevity.

\begin{figure}[htb]
\begin{center}
\epsfxsize=1in\leavevmode\epsfbox{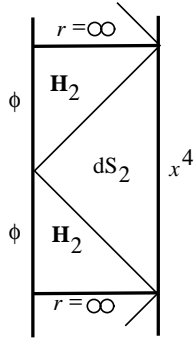}
\caption{dS$_4$ fibered by azimuthal ${\bf H}_2$, dS$_2$.  These
vertical cards have been compactified.}
\label{dScard2fig}
\end{center}
\end{figure}

We give one more example of an interesting 4d de Sitter card diagram: the
purely time-dependent one where dS$_4$ is fibered by azimuthal dS$_2$ and ${\bf H}_2$;
see Fig.~\ref{dScard2fig}.  Each 45-45-90 triangle with a vertical hypotenuse
is a compactified representation of a
half-plane vertical card without special null lines.  The cards are compactified
in precisely the same way as for a Penrose diagram.
The S-RNAdS$_4$ solution asymptotes
down through the ${\bf H}_2$-fibered region
to the two $r=\infty$ regions drawn, with the exception of the `point' on the
right side (actually a $\phi$-circle) where the $x^4$-circle would vanish.

Lastly we look at 5d case and find a card diagram which has both 5d de Sitter
and anti-de Sitter in the same diagram.  Take dS$_5$ fibered by $S^3$ with
$d\Omega_3^2=d\theta^2+\cos^2\theta d\phi^2+\sin^2\theta d\psi^2$;
it has ${dr^2\over 1-r^2/l^2}+r^2d\theta^2$ just like the 4d case.  We will
get an elliptic card diagram.
We want the rightward $z$-ray boundary to be $\theta=0$ and now we want the leftward $z$-ray boundary
to be $\theta=\pi/2$, not $\pi$.  So in analogy with 4d RN,
we need the metric to look like
${dv^2\over v^2-v_0^2}+4d\theta^2$.

To this end, let $u=1/r^2$ and then $u=v+1/2l^2$.  The metric is
$\propto {dv^2\over v^2-1/4l^4}+4d\theta^2$ and we let $v=(1/2l^2)\cosh\zeta$.
The card diagram is then as follows: take the card diagram structure of
the Schwarzschild black hole \cite{Jones:2004pz}; call the positive-mass
external universe the `primary' horizontal card and the negative-mass universe the
`secondary' horizontal card.  Alternatively, take Fig.~\ref{dScard1fig} and
label the alternating levels of horizontal cards primary and secondary.
Two primary horizontal cards and two vertical cards which connect at an
$r=l$ horizon form a dS$_5$ of signature $++++-$.  Each secondary horizontal card forms an AdS$_5$ of opposite signature $+----$.  The horizontal edges of the vertical cards
which are not horizons give $r=\infty$ regions for the dS$_5$ and AdS$_5$
universes that `meet' there.
Note that both $++++-$ dS$_5$ and $----+$ AdS$_5$ satisfy the
Einstein-$\Lambda$ equation with the same $\Lambda>0$, and hence
can appear on the same card diagram.

These diagrams also apply to orbifolds of pure (A)dS space, such as the
constant-curvature black hole of \cite{Cai:2002mr}, which has a non-Killing
horizon, and bears certain resemblance to a BHAdS bubble.

A more complete treatment of generalized card diagrams may appear elsewhere.

\end{document}